\documentclass[11pt]{article}

\usepackage[a4paper,margin=1in]{geometry}
\usepackage[T1]{fontenc}
\usepackage{lmodern}
\usepackage{microtype}
\usepackage{setspace}
\usepackage{graphicx}
\graphicspath{{figures/}}
\usepackage{float}
\usepackage[font=small,labelfont=bf,labelsep=period,justification=centering]{caption}

\usepackage{booktabs}
\usepackage{array}
\usepackage{tabularx}
\usepackage{longtable}
\newcolumntype{L}[1]{>{\raggedright\arraybackslash}p{#1}}

\usepackage{enumitem}
\setlist{itemsep=2pt,topsep=4pt}
\usepackage{amsmath,amssymb}

\usepackage[explicit]{titlesec}
\titleformat{\section}{\Large\bfseries}{\thesection}{0.6em}{#1}
\titleformat{\subsection}{\large\bfseries}{\thesubsection}{0.5em}{#1}
\titleformat{\subsubsection}{\normalsize\bfseries}{\thesubsubsection}{0.4em}{#1}
\titlespacing*{\section}{0pt}{1.4\baselineskip}{0.6\baselineskip}
\titlespacing*{\subsection}{0pt}{1.0\baselineskip}{0.4\baselineskip}
\titlespacing*{\subsubsection}{0pt}{0.8\baselineskip}{0.3\baselineskip}

\usepackage[hidelinks,breaklinks=true,colorlinks=false]{hyperref}
\usepackage{url}
\renewenvironment{abstract}
  {\begin{center}\bfseries\large Abstract\end{center}%
   \begin{quotation}\small\noindent\ignorespaces}
  {\end{quotation}}

\title{\bfseries AMD SEV-SNP: A Confidential Computing Primer}
\author{%
  \begin{tabular}{c@{\hskip 1.5em}c@{\hskip 1.5em}c}
    Amean Asad & Patrick McClurg & Patrick Woodhead
  \end{tabular}
  \\[1.4em]
  \href{https://confidential.ai}{Confidential.ai}
}
\date{June 2026}

\begin{document}
\maketitle

\begin{abstract}
This paper is a technical primer on AMD Secure Encrypted Virtualization with Secure Nested Paging (SEV-SNP), a hardware confidential computing implementation that provides Trusted Execution Environments (TEEs) for virtual machines. SEV-SNP treats the hypervisor as adversarial. It encrypts guest memory and register state with keys the hypervisor never possesses, detects any tampering with guest memory at the point of access, and lets a guest prove to a remote verifier exactly what code it is running. The paper constructs each of these guarantees from the hardware up. It opens with the threat model that drives the design and the hardware that enforces it, the AMD Secure Processor and the encryption engine in the memory controller. It then develops the mechanisms that make a confidential guest practical. The Reverse Map Table provides memory integrity against an adversary who controls the page tables. The privilege and communication machinery (VM Privilege Levels, the encrypted VM Save Area, and the GHCB protocol) lets the guest cooperate with a hypervisor it does not trust. The attestation pipeline binds a hardware-signed measurement of the guest's initial state to AMD's certificate chain, so a remote verifier can confirm independently what is running.

\end{abstract}
\section{Introduction}
\label{sec:introduction}

\subsection{Motivation}
\label{subsec:motivation}

Confidential computing aims to protect data while it is being processed, closing the gap left by encryption at rest and encryption in transit. The dominant deployment substrate for such workloads is the public cloud, where a tenant's virtual machine runs on hardware managed by another party. In conventional virtualization, the hypervisor has complete read and write access to the guest's memory and registers. Any party with sufficient privilege on the host (a cloud operator, a compromised hypervisor, a privileged insider) can observe or tamper with the guest's execution. Cloud providers respond with operational controls (background checks, access audits, network segmentation), but those controls are administrative rather than cryptographic. They do not change the underlying architecture.

Hardware TEEs change the architecture. They move the confidentiality and integrity boundary from a contractual perimeter enforced by the operator's policies and processes to a cryptographic perimeter enforced by silicon. Inside the boundary, memory is encrypted with keys the hypervisor never possesses, and tampering with guest state is detected by hardware checks performed on every relevant memory access. Outside the boundary, the hypervisor still schedules the guest, services its I/O, and manages the platform, but it cannot read or silently modify what the guest computes.

AMD SEV-SNP is one such design. It builds on more than a decade of incremental work on x86 virtualization and memory encryption, culminating in a model that treats the hypervisor as adversarial and supplies hardware-enforced confidentiality, integrity, and remote attestation for an entire virtual machine.

\subsection{Scope and Audience}
\label{subsec:scope-and-audience}

This paper provides a technical primer on SEV-SNP. It is intended as a single document that a reader can consult to understand what the technology guarantees, how those guarantees are implemented, and how the resulting attestations are verified in practice. The treatment proceeds from threat model to hardware foundations, then to integrity, communication, and attestation. Each section is written so that a reader who has the prerequisites for the section before it can follow without prior exposure to the rest of the document.

The intended reader is comfortable with basic virtualization concepts (virtual machines, hypervisors, nested paging) and with public-key cryptography at the level of signatures and certificate chains. No prior knowledge of AMD-specific architecture is assumed. The paper does not attempt to be a substitute for AMD's specifications; where exact byte offsets, instruction encodings, or register field layouts matter for an implementation, the relevant AMD publications \cite{ref1,ref2,ref3} should be consulted directly.

The treatment is deliberately specific to SEV-SNP rather than couched in fully generic TEE language. The conceptual machinery (a hardware root of trust, a measured launch, an attestation report, a verifier that checks measurement and policy against expected values) generalizes to Intel TDX \cite{ref4}, Arm CCA \cite{ref5}, and other current and forthcoming CPU TEE designs. The IETF RATS architecture \cite{ref6} frames the verifier's role and the attestation evidence flow in vendor-neutral terms, and an overview of trusted execution environments as a discipline appears in Jauernig et al. \cite{ref7}. Concrete implementation details (the Reverse Map Table, the C-bit, the VMPL hierarchy, the VCEK key derivation) are SEV-SNP-specific. The reader interested in cross-vendor comparison can map these mechanisms onto their analogues in other architectures, which mostly differ in naming and in details of the firmware-level interface rather than in fundamental structure.

\section{Threat Model}
\label{sec:threat-model}

The threat model is the design driver. SEV-SNP exists to enforce specific guarantees against a specific adversary; the rest of the architecture follows from that specification. This section states the guarantees, names the parties on each side of the trust boundary, and is explicit about what falls outside the boundary entirely.

\subsection{Security Guarantees}
\label{subsec:security-guarantees}

SEV-SNP provides three security properties for a confidential virtual machine.

\textbf{Confidentiality.} Guest memory is encrypted with a per-VM key that the hypervisor cannot access. Register state is encrypted when the VM exits to the hypervisor. The hypervisor sees ciphertext, not plaintext; on current parts the guest can additionally require ciphertext hiding, which denies the hypervisor visibility of even the ciphertext (Section 3.3). This is a cryptographic guarantee enforced by the on-die memory controller and the AMD Secure Processor; defeating it requires breaking the memory-encryption cipher (AES-128 on Milan parts, AES-256 on Genoa and later) or compromising AMD's key management.

\textbf{Integrity.} The hardware detects when the hypervisor tampers with guest memory. Substituting pages, replaying old data, and remapping addresses each trigger faults the guest can catch. The hypervisor cannot silently corrupt guest state. This too is a hardware-enforced guarantee, implemented by checks performed on every relevant memory access (Section 4).

\textbf{Attestation.} The guest can prove to a remote party what code it is running. The AMD Secure Processor measures the initial guest image and signs a report whose signing key chains to fuses set during chip manufacturing. A verifier checks the signature against AMD's certificate chain to confirm that the report originated from genuine AMD hardware and not from software pretending to be a confidential VM (Section 6).

SEV-SNP does not guarantee availability. The hypervisor controls scheduling, memory provisioning, and physical resources, and may terminate a guest at any time. Availability cannot be enforced against an adversary who controls the power switch, and the design does not attempt to. Workloads that need availability guarantees obtain them through operational means (redundancy, multi-cloud, service-level agreements) rather than from the TEE itself.

\subsection{The Adversarial Hypervisor}
\label{subsec:the-adversarial-hypervisor}

The defining design choice of SEV-SNP is that it treats the hypervisor as adversarial. This is not a statement about the intentions of any particular cloud provider; it is a worst-case specification. The hardware is designed so that a compromised or malicious hypervisor cannot read guest memory, register state, or silently modify guest data without detection.

A fully privileged hypervisor adversary in this context can do all of the following.

\begin{itemize}
  \item Read and write any region of host memory.
  \item Inspect and modify the nested page tables that translate guest physical addresses to system physical addresses.
  \item Decide which physical pages back the guest's memory.
  \item Intercept and inject interrupts.
  \item Pause, resume, snapshot, and migrate the guest at will.
  \item Control the I/O devices the guest believes it is interacting with.
\end{itemize}

In a conventional virtualization stack, this level of access is total compromise. Memory can be dumped, keys extracted from RAM, code modified in flight, and every byte the guest processes observed. SEV-SNP's goal is to make most of these attacks ineffective despite the adversary having this level of access. The mechanisms that achieve this (memory encryption, the Reverse Map Table, encrypted register save areas, the GHCB protocol) are the subject of Sections 3 through 5.

\subsection{Trust and Distrust}
\label{subsec:trust-and-distrust}

The value of a TEE comes not from what it protects, but from what no longer needs to be trusted. The trust boundary defined by SEV-SNP is narrow and explicit.

\textbf{Trusted.} The model trusts AMD. Specifically, it trusts that AMD's manufacturing process did not embed backdoors in the silicon, that AMD's key management practices keep the root keys secure, that the AMD Secure Processor (ASP) firmware does not have exploitable vulnerabilities, and that AMD's key distribution service serves authentic certificates. It also trusts the cryptographic primitives the design relies on (AES for memory encryption, ECDSA P-384 for attestation signatures, SHA-384 for measurements) and the silicon vendor's microcode and ASP firmware as the root of execution.

The model also trusts the party that physically hosts the running hardware to refrain from invasive physical attacks against the silicon. Memory bus interposition, voltage glitching, chip decapsulation, and similar attacks require physical possession of the running machine and specialized equipment, and they can break TEE guarantees. In current cloud deployments the physical hardware host and the hypervisor operator are typically the same company; the model trusts that company in its physical-host role while explicitly distrusting it in its hypervisor-operator role. Recent results such as TEE.Fail, Battering RAM, and BadRAM \cite{ref8,ref9,ref10} have demonstrated practical interposition and aliasing attacks against current TEE designs; these confirm rather than refute the placement of physical attacks outside the threat model. Even so, AMD has begun narrowing the category. Current firmware performs a boot-time DRAM alias check whose completion is reported in attestation (Section 6.2), a direct response to BadRAM-class aliasing, and AMD's advisories continue to track the interposition and fault-injection classes \cite{ref11,ref12}.

Finally, the model trusts the workload's own code. SEV-SNP protects the execution environment, not the program running inside it. If the guest kernel has a vulnerability, an attacker who can deliver malicious input through legitimate channels can exploit it. Attestation proves what code was loaded, not that the code is correct.

\textbf{Untrusted.} Nothing else in the cloud stack is trusted. The hypervisor, host operating system, management plane, orchestration systems, and monitoring agents all sit outside the trust boundary. Cloud operator employees with root access to the hypervisor and datacenter technicians with physical access to non-running hardware are outside the boundary. Other tenants are outside the boundary; even a compromised peer VM that escalates to hypervisor privilege cannot reach the protected guest's memory. Network and storage infrastructure carry data outside the encrypted memory boundary and are also untrusted; data leaving guest memory is protected by software encryption (TLS, LUKS) rather than by the TEE itself.

\subsection{The Trust Boundary}
\label{subsec:the-trust-boundary}

Putting confidentiality, integrity, and the trust-and-distrust list together, the SEV-SNP trust boundary cuts through the system as follows. The CPU die (cores, caches, the on-die memory controller, and the AMD Secure Processor) is trusted. Everything outside the package, including DRAM, the memory bus, devices attached over PCIe, and all software running on the x86 cores other than the measured guest, is treated as untrusted.

\begin{figure}[H]
\centering
\includegraphics[width=0.92\linewidth]{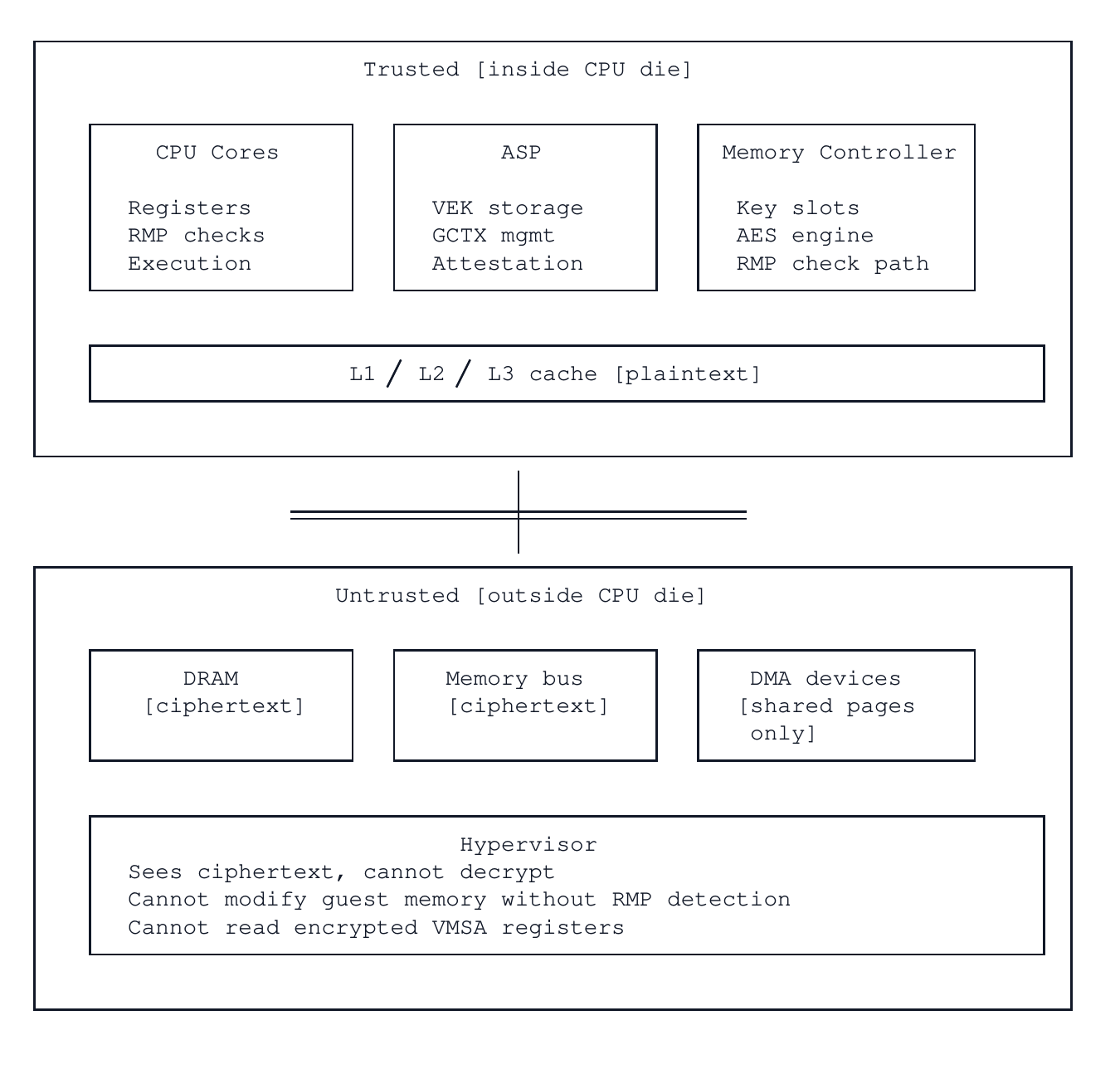}
\caption{The SEV-SNP trust boundary. The on-die memory controller encrypts and decrypts data crossing the die boundary; the AMD Secure Processor (ASP) manages keys; the cores enforce RMP checks inline with memory access. The hypervisor and all software outside the measured guest sit on the untrusted side.}
\label{fig:01-trust-boundary}
\end{figure}

\subsection{Out of Scope}
\label{subsec:out-of-scope}

Three categories of attack are explicitly outside the SEV-SNP threat model. They are architectural exclusions rather than gaps to be closed in a future version.

\textbf{Availability and resource control.} The hypervisor schedules the guest, allocates physical memory, and controls I/O bandwidth. It can refuse to run the guest, terminate it, or starve it of resources. Timing-sensitive workloads remain vulnerable to interference. Availability guarantees come from operational mechanisms, not from the TEE.

\textbf{Side channels and timing leakage.} The CPU shares microarchitectural state between the guest and the hypervisor: caches, branch predictors, prefetchers, and execution-port contention. AMD has added mitigations for several specific attacks in this class (separate branch prediction state, cache partitioning options) and the bar is significantly higher than for unprotected VMs, but side-channel risk is not eliminated. One member of this class deserves specific mention because hardware has since closed it: the ciphertext side channel, in which the hypervisor infers guest data patterns by observing encrypted memory (exploiting the determinism of the memory-encryption tweak), enabled practical attacks against SEV deployments \cite{ref13,ref14,ref15}. On current parts, ciphertext hiding removes this channel at the memory controller (Section 3.3); it remains a consideration on older parts and wherever the feature is not enabled. The hypervisor also controls when interrupts are delivered and can observe the timing of guest operations even when their content is encrypted; this is itself a side channel.

\textbf{Software vulnerabilities and compromised images.} SEV-SNP protects the execution environment, not the software. A buffer overflow in the guest kernel is exploitable in the same way it would be on any other Linux system, and attestation will faithfully report the measurement of a backdoored image without telling the verifier the image is backdoored. The verifier needs to know which measurements correspond to trustworthy software; the measurement itself is identity, not behavior.

\textbf{Invasive physical attacks.} Memory bus interposition, DRAM probing, voltage glitching, chip decapsulation, and similar attacks require physical possession of the running machine and specialized equipment. They can break TEE confidentiality or integrity. They are outside the software-adversary model the design targets and require a different (and substantially more expensive) defense posture, generally provided operationally by the physical hardware host.

\textbf{Trust anchor compromise.} If AMD's manufacturing process is compromised, or if ASP firmware contains a vulnerability that allows the boundary to be bypassed, the guarantees fail. This is a much smaller attack surface than trusting the entire cloud stack, but it is not zero. A SEV-SNP deployment trades trust in many parties for concentrated trust in one. Nor is it merely hypothetical. Researchers have demonstrated extraction of the root attestation seed on first-generation SNP parts by chaining firmware vulnerabilities, enabling forged attestation reports on those parts \cite{ref16} (Section 6.3).

\section{Hardware Foundations}
\label{sec:hardware-foundations}

SEV-SNP is built on top of AMD's existing virtualization architecture, AMD-V, augmented by the on-die AMD Secure Processor and an AES engine integrated into the memory controller. This section describes those building blocks. It then traces the four-generation evolution from Secure Memory Encryption through SEV, SEV-ES, and SEV-SNP, showing what gap each generation closed.

\subsection{AMD-V Virtualization}
\label{subsec:amd-v-virtualization}

SEV-SNP does not replace virtualization; it adds security properties on top of it. Understanding the underlying virtualization machinery is therefore necessary before the security mechanisms make sense.

The classical problem of virtualization is that an operating system expects to control the entire machine: memory, interrupts, privileged instructions. Running two operating systems side by side without mediation, both expecting that control, is unstable. Early hypervisors solved this through software trapping, intercepting every privileged operation, simulating it, and returning control to the guest. AMD-V (introduced in 2006) and Intel's analogous VT-x added explicit CPU modes for running a guest and running the hypervisor, letting hardware handle most of the isolation directly.

\subsubsection{The VMCB}

The Virtual Machine Control Block (VMCB) is the data structure through which the hypervisor controls a guest. It has two parts: a control area that the hypervisor reads and writes freely, and a save area that holds the guest's CPU state across exits.

\begin{figure}[H]
\centering
\includegraphics[width=0.92\linewidth]{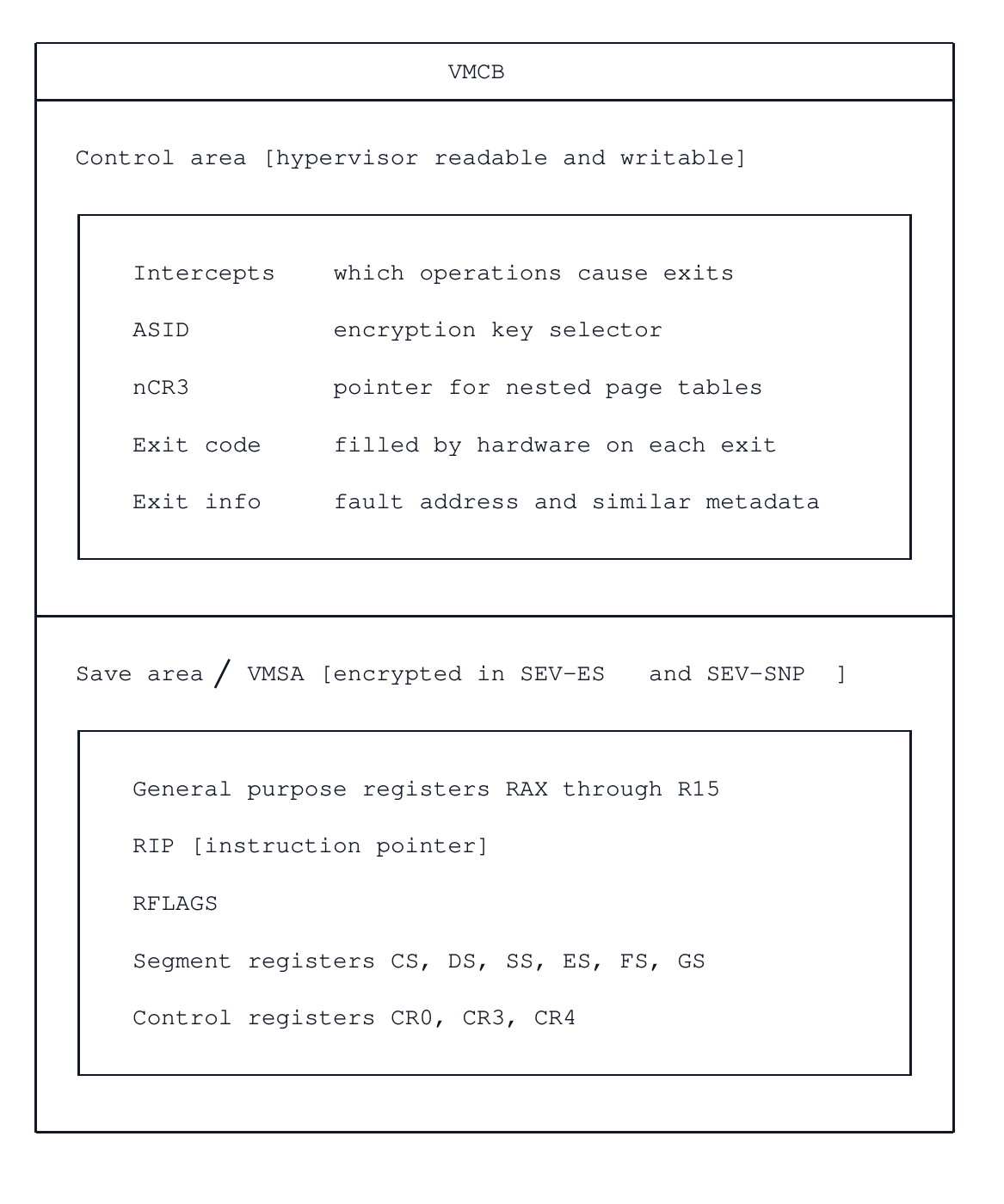}
\caption{The VMCB. The control area carries instructions to the hardware about which operations should cause exits, the ASID that selects the encryption key, the nested page table root, and fields the hardware fills in when an exit occurs. The save area, designated VMSA in SEV-ES and SEV-SNP, holds the guest's full register state across exits and is encrypted in those configurations with the guest's per-VM key.}
\label{fig:02-vmcb-structure}
\end{figure}

The save area distinction matters for security. In plain virtualization, the hypervisor reads and writes the save area directly, which means it sees every register value the guest holds at the moment of exit. SEV-ES makes the save area opaque by encrypting it with the guest's key, and SEV-SNP further marks it as VMSA in the integrity table so the hypervisor cannot write to it either. The save area is the channel through which CPU state would otherwise leak across the trust boundary; closing it is the work of the next generation of the architecture.

\subsubsection{The VMRUN cycle}

Guest execution proceeds in a loop. The hypervisor sets up a VMCB, executes the VMRUN instruction, and the hardware switches to guest mode and resumes the guest at the saved instruction pointer. The guest runs at near-native speed until something triggers an exit: an intercepted instruction, an I/O access, a page fault, or an external interrupt. At that point the hardware saves the guest's registers back to the save area, writes exit information into the control area, restores hypervisor state, and returns control to the hypervisor at the instruction following VMRUN. The hypervisor reads the exit code, services whatever caused the exit, and calls VMRUN again.

\begin{figure}[H]
\centering
\includegraphics[width=0.95\linewidth]{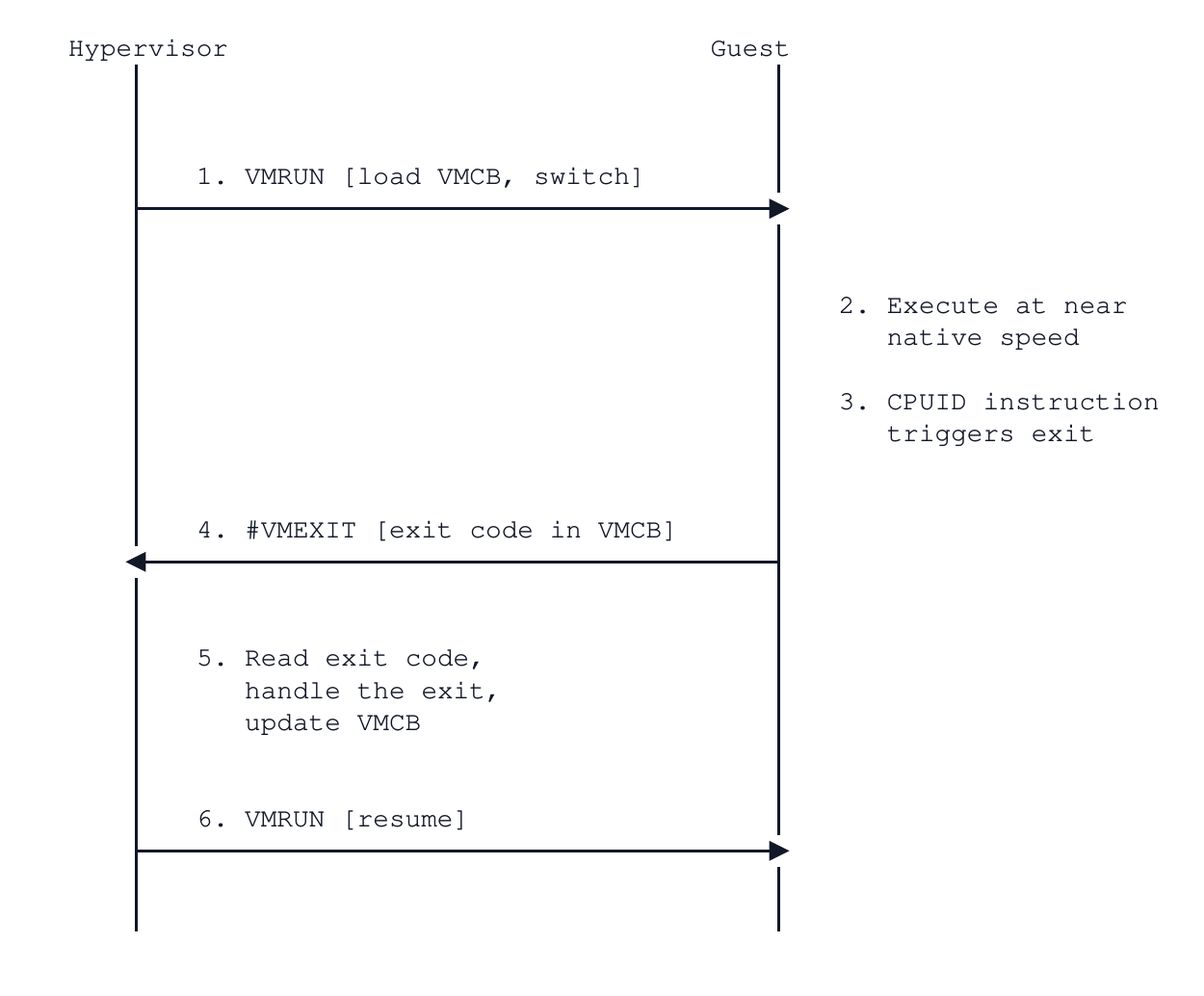}
\caption{The VMRUN cycle. The hypervisor and guest alternate, with the hypervisor seeing only the exit code and metadata it has been told to capture. Most guest execution happens at full speed; the hypervisor is involved only when an intercepted operation occurs.}
\label{fig:03-vmrun-cycle}
\end{figure}

This cycle repeats millions of times per second. In SEV-SNP, the same cycle still drives execution, but the hypervisor's view of what happens during a guest run is sharply restricted. It still observes the fact of an exit and the exit code that hardware writes to the control area, but the register state that classical virtualization would have made visible is encrypted and inaccessible.

\subsubsection{Nested paging}

Before hardware nested paging, hypervisors maintained \emph{shadow page tables}, a parallel structure mapping guest virtual addresses directly to system physical addresses, kept in sync with the guest's own tables by intercepting every guest page-table modification. The performance cost was large.

AMD-V introduced nested paging (Intel's analogue is EPT, Extended Page Tables). The CPU now performs two levels of translation natively. The guest's page tables, controlled by the guest OS, translate a guest virtual address (GVA) to a guest physical address (GPA). The nested page tables, controlled by the hypervisor, translate the GPA to the system physical address (SPA) where the data actually lives in DRAM.

\begin{figure}[H]
\centering
\includegraphics[width=0.7\linewidth]{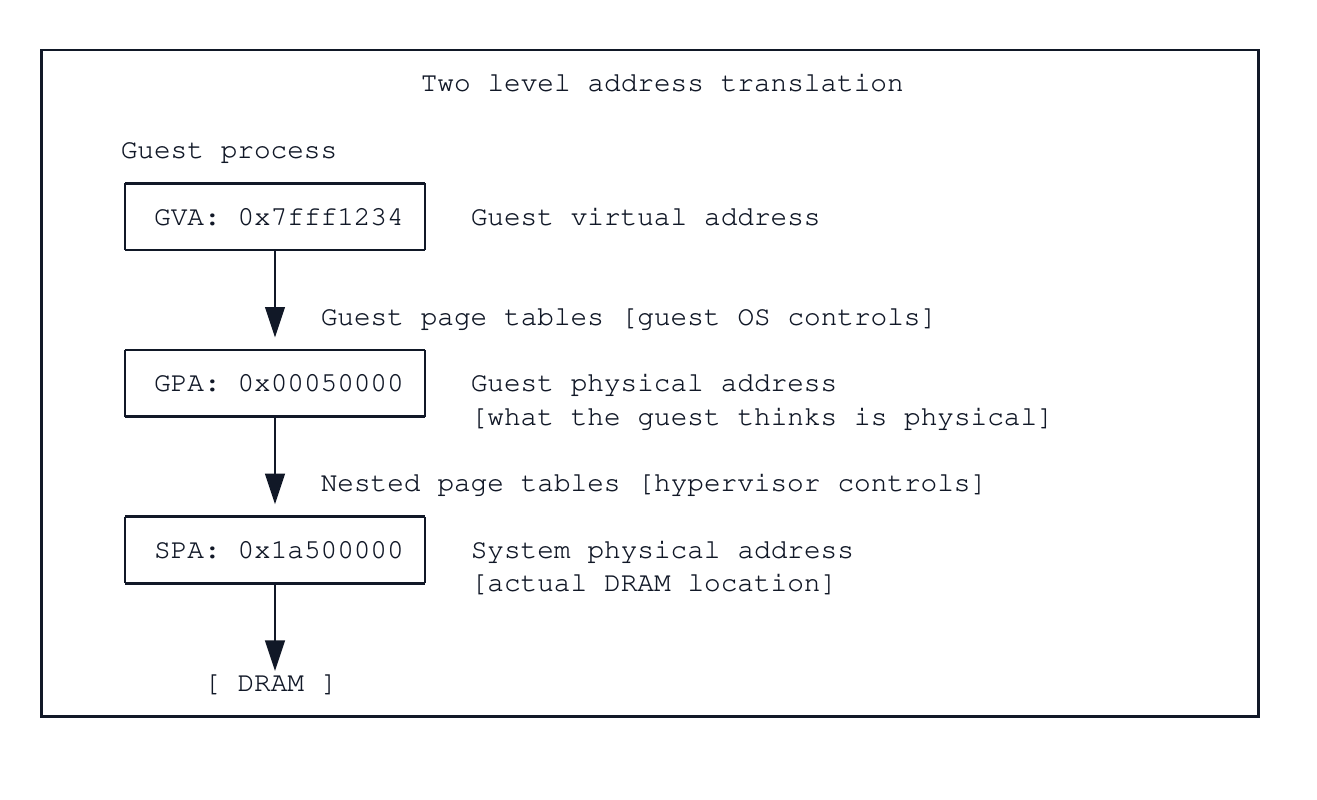}
\caption{Two-level address translation under nested paging. The hypervisor controls the GPA-to-SPA step, and that control is the lever a malicious hypervisor would use to remap, alias, or replay guest memory. SEV-SNP closes this lever with the integrity check described in Section 4.}
\label{fig:04-nested-paging}
\end{figure}

The nested page table root is stored in the VMCB's nCR3 field. Guest page-table modifications no longer cause exits; the hypervisor is invoked only when GPA-to-SPA translation fails, indicating that the guest needs more memory.

The hypervisor's control over the nested page tables is necessary for virtualization but is also the source of the integrity attacks that SEV-SNP addresses. A malicious hypervisor could remap a GPA to a different SPA, alias multiple GPAs to the same SPA, or replay an old SPA's contents at a new time. Encryption alone does not prevent any of these, because they manipulate which ciphertext the guest sees rather than the ciphertext itself. The Reverse Map Table (Section 4) closes this gap.

\subsection{The AMD Secure Processor}
\label{subsec:the-amd-secure-processor}

The AMD Secure Processor (ASP; formerly the Platform Security Processor, PSP, a name still common in the SEV literature) is a dedicated ARM Cortex-A5 microcontroller integrated onto the AMD CPU die. It is part of the silicon, not a separable component, and it boots before the x86 cores; on an AMD EPYC system the ASP initializes the platform and only then does the x86 instruction stream begin. The ASP runs its own firmware, independent of the x86 cores, and holds the keys that the rest of the architecture treats as cryptographic ground truth.

The ASP's responsibilities span four areas. First, it generates per-VM encryption keys, called VEKs (VM Encryption Keys). Each is an AES key (128-bit on Milan parts, 256-bit on Genoa and later) produced by an on-chip NIST SP 800-90 hardware random number generator, unique per guest instance even when two guests boot identical images. Second, it maintains a Guest Context (GCTX) page for each SNP guest, holding the VEK, measurement digest, policy, communication keys (the VMPCKs introduced in Section 5), and other per-guest security state; this page is encrypted by the ASP and marked immutable in the integrity table, so the hypervisor sees only ciphertext and cannot modify the entry. Third, it owns the VM lifecycle: launch, activation, deactivation, and migration commands all flow through the ASP, which performs the operation and enforces security invariants. Fourth, it generates and signs attestation reports, using a key derived from fuses set during chip manufacturing.

The x86 cores communicate with the ASP through MMIO mailbox registers in PCI configuration space, using a request-response protocol. The x86 software allocates a command buffer in DRAM, writes the buffer's physical address to mailbox registers, writes a command identifier, and waits for the ASP to consume the command, perform the operation, and signal completion. The interface is intentionally coarse-grained. The ASP does not sit in the fast path of guest execution, and it handles security-critical operations only.

The architectural significance of having the ASP as a separate processor with its own firmware is that the x86 hypervisor cannot read ASP internal state, regardless of what privilege it holds on the x86 side. VEKs never leave the ASP in cleartext. They are programmed into the memory controller's key slots over a hardware interface that the hypervisor does not have access to. Even an attacker with full hypervisor control cannot extract a VM's encryption key.

\subsection{Memory Encryption}
\label{subsec:memory-encryption}

The encryption itself happens in dedicated AES engines located in the on-die memory controllers, not in the CPU cores. Encryption and decryption occur at the boundary between the CPU package and DRAM: data inside the CPU caches is plaintext, and data leaving the package toward DRAM is ciphertext. Encrypting on every cache hit would be unacceptably slow, and the threat model treats the CPU package as trusted, so the cache boundary is the right place.

\begin{figure}[H]
\centering
\includegraphics[width=0.92\linewidth]{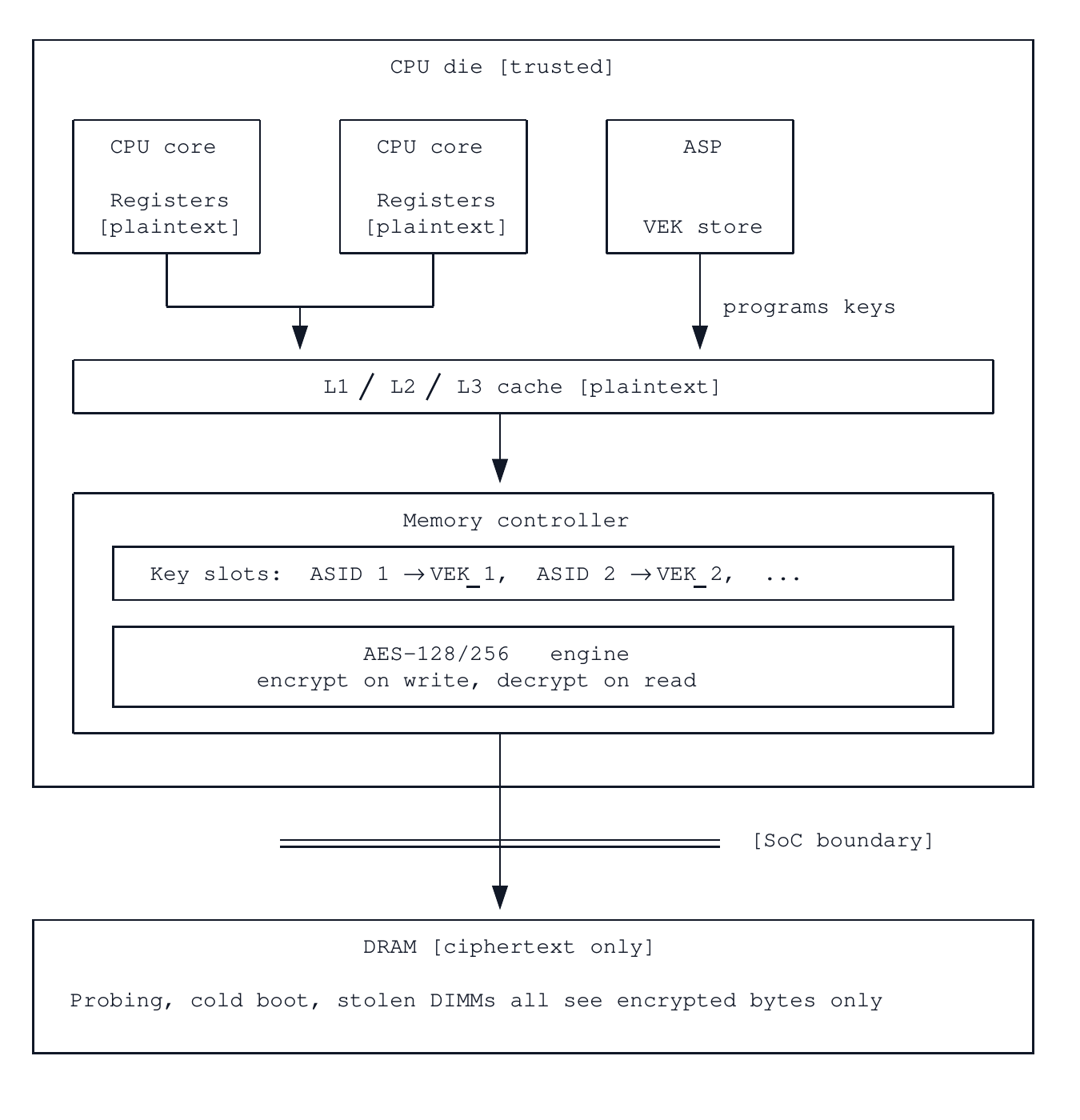}
\caption{The memory encryption data path. Plaintext lives inside the package (registers, caches); ciphertext lives in DRAM. The memory controller holds a key slot per ASID and selects the appropriate VEK using the ASID tag carried with each memory transaction.}
\label{fig:05-memory-encryption}
\end{figure}

The memory controller maintains key slots, one per ASID. When a memory transaction arrives, it carries the ASID of the originating guest (set in the VMCB and propagated through the memory pipeline by hardware). The memory controller looks up the VEK for that ASID and uses it to encrypt outgoing writes and decrypt incoming reads. The encryption uses AES with the physical address as a tweak, so the same plaintext at different physical addresses encrypts to different ciphertext. AMD deploys this in two modes tied to key size: AES-128-XEX on Milan parts and AES-256-XTS on Genoa and later (XTS is the standardized construction built on the XEX tweak). Which mode a guest gets is governed by launch policy bit 22, MEM\_AES\_256\_XTS. When it is cleared the platform may use either AES-128-XEX or AES-256-XTS, and when it is set the guest requires AES-256-XTS. This blocks ciphertext block-move attacks where an attacker copies encrypted data from one location to another.

The tweak is deterministic, however. The same plaintext rewritten to the same address produces the same ciphertext, and because the hypervisor could always observe guest ciphertext, that determinism enabled a class of ciphertext side-channel attacks that infer guest data from memory write patterns \cite{ref13,ref14}. AMD's answer on current parts is ciphertext hiding. When the platform enables it, the memory controller stops exposing guest ciphertext to non-owner accesses, so the hypervisor no longer has ciphertext to analyze \cite{ref15}. A guest can require ciphertext hiding in its launch policy, and whether it is active is visible in the attestation report (Section 6.2); on older parts, or wherever the feature is not enabled, the ciphertext side channel remains a consideration.

\subsubsection{The ASID}

The ASID (Address Space Identifier) is an identifier that tags memory transactions, carried as a 32-bit field in the SEV API and SNP ABI command buffers. The hypervisor sets the ASID in the VMCB before executing VMRUN, and hardware ensures all memory accesses originating from the guest carry that ASID through the cache hierarchy and into the memory controller. The number of ASIDs available on a given processor is exposed through CPUID and is typically a few hundred. When the workload requires more concurrent confidential VMs than there are ASIDs, the hypervisor must overcommit: deactivate one guest, flush caches, and activate another, with the SNP\_ACTIVATE and SNP\_DEACTIVATE ASP commands installing and removing VEKs from key slots.

\begin{figure}[H]
\centering
\includegraphics[width=0.92\linewidth]{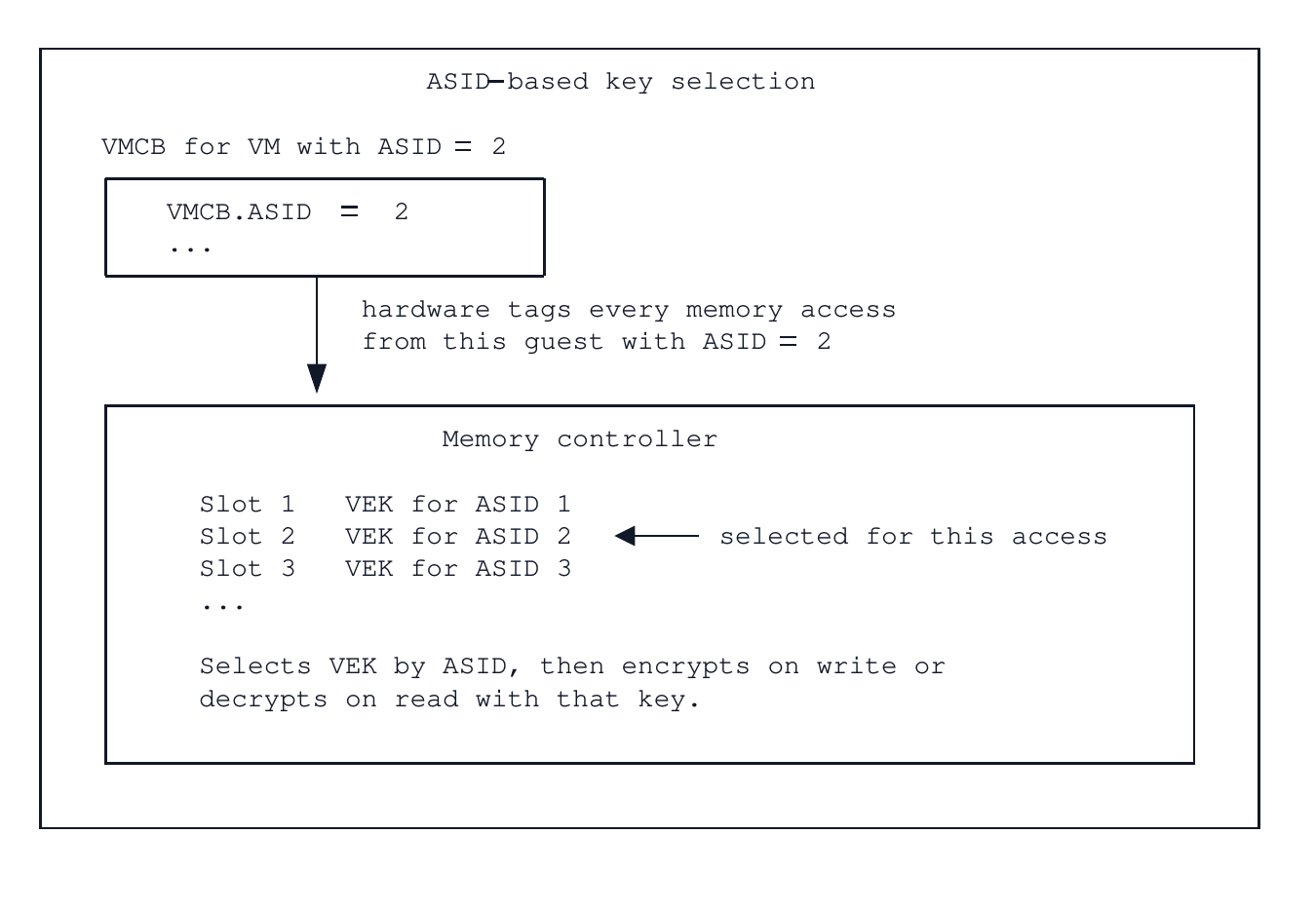}
\caption{ASID-based key selection. Every memory access from a guest carries the guest's ASID; the memory controller uses that ASID to index into a small bank of VEK slots and selects the appropriate key for encryption or decryption. Different ASIDs select different keys, so memory belonging to one guest cannot be decrypted under another guest's key.}
\label{fig:06-asid-key-selection}
\end{figure}

\subsubsection{The C-bit}

Within a guest, individual pages can be marked as either private (encrypted with the guest's VEK) or shared (not encrypted with the guest's VEK). The selector is the C-bit, a single bit of guest physical addresses in page table entries (its exact position is processor-dependent and reported via CPUID). Private pages (C=1) are the default for guest memory. Shared pages (C=0) are used for explicit communication with the hypervisor (the GHCB, Section 5.3) and for DMA buffers, since devices performing DMA do not execute in guest context and therefore have no ASID to select a key.

The guest sets the C-bit in its own page tables. The hypervisor cannot override this choice. For I/O the guest typically uses bounce buffers. A shared buffer is allocated, data copied into it, the device performs DMA, and the result is copied back into private memory. Linux's SWIOTLB implements this transparently for SEV guests. Trusted I/O (Section 4.6) relaxes this for TDISP-capable devices on current parts: a device bound into the guest's trust domain can DMA directly into private memory, and no shared-page bounce is needed on that path.

\subsection{The Generational Evolution}
\label{subsec:the-generational-evolution}

SEV-SNP is the fourth in a sequence of memory-encryption features, each closing gaps left by its predecessor. The progression clarifies what protections SNP actually provides and which earlier limitations it removes.

\textbf{SME (Secure Memory Encryption).} SME introduced the AES engine in the memory controller and the C-bit \cite{ref17}. It uses a single system-wide key for all encrypted memory; the OS or hypervisor controls which pages are encrypted. This protects against physical attacks (cold boot, memory bus probing, stolen DIMMs) but not against any software adversary on the host, since everyone with software access to memory can decrypt by simply reading.

\textbf{SEV (Secure Encrypted Virtualization).} SEV introduced per-VM keys. Each VM gets its own VEK, selected by ASID, and the hypervisor reading guest memory now sees ciphertext it cannot decrypt. This protects guest memory contents from a software-level hypervisor adversary. It does not protect register state. On every VMEXIT the guest's registers are saved into the VMCB save area in plaintext, where the hypervisor can read or modify them at will.

\textbf{SEV-ES (Encrypted State).} SEV-ES closes the register exposure gap. The save area, now called the VMSA, is encrypted with the guest's VEK on every exit. The hypervisor sees ciphertext where it previously saw register values. Because the hypervisor can no longer see what the guest needs when an exit occurs, SEV-ES introduces a new mechanism for guest-hypervisor cooperation: the \#VC exception, raised inside the guest when an intercepted operation occurs, and the GHCB protocol through which the guest places explicit information in a shared page for the hypervisor to consume (Section 5). What SEV-ES still does not protect against is memory integrity. A malicious hypervisor can replace a page with an old copy, flip bits in encrypted memory, alias multiple GPAs to the same SPA, or silently remap which physical page backs a GPA. None of these allow the attacker to read data, but each can corrupt or stale the data the guest operates on. Practical attacks against the SEV and SEV-ES generations demonstrated each of these failure modes, including memory remapping (SEVered \cite{ref18}), ciphertext malleability (SEVurity \cite{ref19}), and fault injection against the ASP itself \cite{ref20}, motivating the integrity work in SEV-SNP.

\textbf{SEV-SNP (Secure Nested Paging).} SEV-SNP adds hardware-enforced integrity. The fundamental guarantee is that if a guest reads a private page, it sees the value it last wrote, or it gets an exception; never stale data, never corrupted data, never another page's data. The mechanism is the Reverse Map Table (Section 4), a system-wide table that records, for every physical page, which VM owns it and at what guest physical address it should appear. Every memory access to a private guest page triggers an RMP check inline with translation; a mismatch causes a fault. SEV-SNP also introduces VM Privilege Levels, allowing privilege separation within an encrypted guest (Section 5.1), and adds restricted interrupt injection, branch-prediction barriers, and TCB versioning of the attestation key derivation.

The protection coverage across the four generations is summarized in the following table:

\begin{table}[H]
\centering
\begin{tabular}{lllll}
\toprule
\textbf{Threat} & \textbf{SME} & \textbf{SEV} & \textbf{SEV-ES} & \textbf{SEV-SNP} \\
\midrule
Some physical DRAM attacks & Yes & Yes & Yes & Yes \\
Hypervisor reads VM memory & No & Yes & Yes & Yes \\
Hypervisor reads VM registers & No & No & Yes & Yes \\
Memory replay & No & No & No & Yes \\
Memory corruption & No & No & No & Yes \\
Memory aliasing & No & No & No & Yes \\
Memory remapping & No & No & No & Yes \\
\bottomrule
\end{tabular}
\end{table}

The threat model laid out in Section 2 corresponds specifically to the SEV-SNP column. Each row reflects a class of attack that the architecture closes; any deployment relying on an earlier generation inherits the gaps of that generation rather than the SNP guarantee.

SNP's introduction was not the end of the hardware's evolution. Subsequent processor generations have layered new capabilities onto the SNP baseline, and current parts add several that matter to the discussions in later sections: ciphertext hiding, which denies the hypervisor visibility of guest ciphertext (Section 3.3); Secure AVIC, which filters interrupt injection in hardware (Section 5.4); a segmented RMP option, which improves multi-socket scalability (Section 4.1); performance-counter virtualization, which removes the PMU as a hypervisor side channel; finer-grained guest intercept controls; a secure TSC, giving the guest a clock the hypervisor cannot skew; and Trusted I/O, which binds TDISP-capable PCIe devices into the guest's trust domain (Section 4.6) \cite{ref1,ref3}. These features are individually optional, enabled by the platform or required by the guest's launch policy. Two of them are visible to a verifier in the attestation report's PLATFORM\_INFO field (ciphertext hiding and Trusted I/O, alongside the boot-time alias check; Section 6.2); the rest are matters of platform and guest configuration that attestation does not directly cover.

\subsection{Key Hardware Structures}
\label{subsec:key-hardware-structures}

The rest of the paper relies on the following hardware structures.

\begin{itemize}
  \item \textbf{ASID key slots.} A small number of slots in the memory controller, each holding one VEK. The ASP programs them via SNP\_ACTIVATE and clears them via SNP\_DEACTIVATE.
  \item \textbf{The RMP.} A system-wide table with one 16-byte entry per 4 KB physical page, indexed by system physical address, recording owner, expected GPA, validation state, and per-VMPL permissions. Modified only by specific privileged instructions or ASP commands. Inline RMP checks are performed by CPU microarchitecture on every relevant memory access (Section 4).
  \item \textbf{The GCTX.} A 4 KB page maintained by the ASP for each guest, holding the VEK, measurement digest, communication keys, and policy. The ASP encrypts it and marks it immutable in the RMP, so the hypervisor sees ciphertext but cannot modify it.
  \item \textbf{The VMSA.} The encrypted save area carrying guest register state. Encrypted with the guest's VEK. The hypervisor specifies a pointer to it in the VMCB control area but cannot read or write its contents.
  \item \textbf{The GHCB.} A shared (C=0) page used for explicit guest-hypervisor communication. Visible to both sides, with the guest controlling what is placed in it (Section 5.3).
\end{itemize}

These structures are introduced individually as the paper proceeds, but they are all rooted in the hardware described in this section: the memory controller's encryption engine, the ASP's key management, and the cores' inline integrity check path.

\section{Memory Integrity}
\label{sec:memory-integrity}

Memory encryption alone protects confidentiality, since the hypervisor sees ciphertext rather than plaintext. It does not protect integrity, because encryption does not prevent the hypervisor from manipulating which ciphertext the guest sees. The hypervisor controls the nested page tables and therefore decides which physical page backs each guest physical address. Even with encryption, four classes of attack remain available:

\begin{itemize}
  \item \textbf{Remapping.} The guest writes sensitive data to GPA 0x1000, which is backed by SPA X. Later, the hypervisor changes the nested page table so that GPA 0x1000 maps to SPA Y. The guest reads GPA 0x1000 expecting its data and instead gets whatever was at SPA Y. No decryption was needed.
  \item \textbf{Replay.} The hypervisor captures the ciphertext at SPA X at time T1. Later, after the guest updates that memory, the hypervisor restores the old ciphertext. The guest sees stale data without realizing it.
  \item \textbf{Aliasing.} The hypervisor maps two different guest addresses (GPA A and GPA B) to the same physical page (SPA X). Writes to GPA A appear at GPA B and vice versa, corrupting any data structure that assumed those addresses were independent.
  \item \textbf{Corruption.} The hypervisor writes random bytes into a guest's physical memory. The ciphertext is garbage and decrypts to garbage. The hypervisor cannot control the resulting plaintext, but corrupting a critical data structure can crash the guest or steer it into exploitable behavior.
\end{itemize}

SEV and SEV-ES had no defense against any of these. SEV-SNP closes the gap with the Reverse Map Table (RMP) and a page-validation discipline enforced by hardware. The combination guarantees that if a guest reads a private page, it sees the value it last wrote, or it gets an exception.

\subsection{The Reverse Map Table}
\label{subsec:the-reverse-map-table}

The RMP is a system-wide table in DRAM with one entry per 4 KB physical page, indexed by system physical address. Each entry answers two questions: who owns this physical page, and at what guest physical address should it appear?

The RMP is configured via dedicated MSRs (RMP\_BASE and RMP\_END) and is 1 MB-aligned. In its original form it is a single table shared by all sockets in a multi-socket system; current parts also support a segmented RMP, in which each node manages its own segment, improving scalability and locality on large multi-socket machines \cite{ref3}. Each 16-byte entry contains:

\begin{table}[H]
\centering
\small
\begin{tabularx}{\linewidth}{@{} >{\bfseries}p{0.28\linewidth} X @{}}
\toprule
Field & \textbf{Purpose} \\
\midrule
Assigned & Is this page assigned to a guest or to firmware? \\
\addlinespace
ASID & Which guest owns this page [0 = hypervisor / free] \\
\addlinespace
GPA & The guest physical address this page should map to \\
\addlinespace
Validated & Has the guest explicitly accepted this page? \\
\addlinespace
Immutable & Is this page locked for firmware operations? \\
\addlinespace
VMSA & Is this a VM Save Area page? \\
\addlinespace
Page\_Size & 4 KB or 2 MB \\
\addlinespace
VMPL perms & Read / write / execute permissions per VMPL [0..3] \\
\bottomrule
\end{tabularx}
\end{table}

The combination of these fields determines the page's state. Three mutable states are encountered most often.

\begin{figure}[H]
\centering
\includegraphics[width=0.92\linewidth]{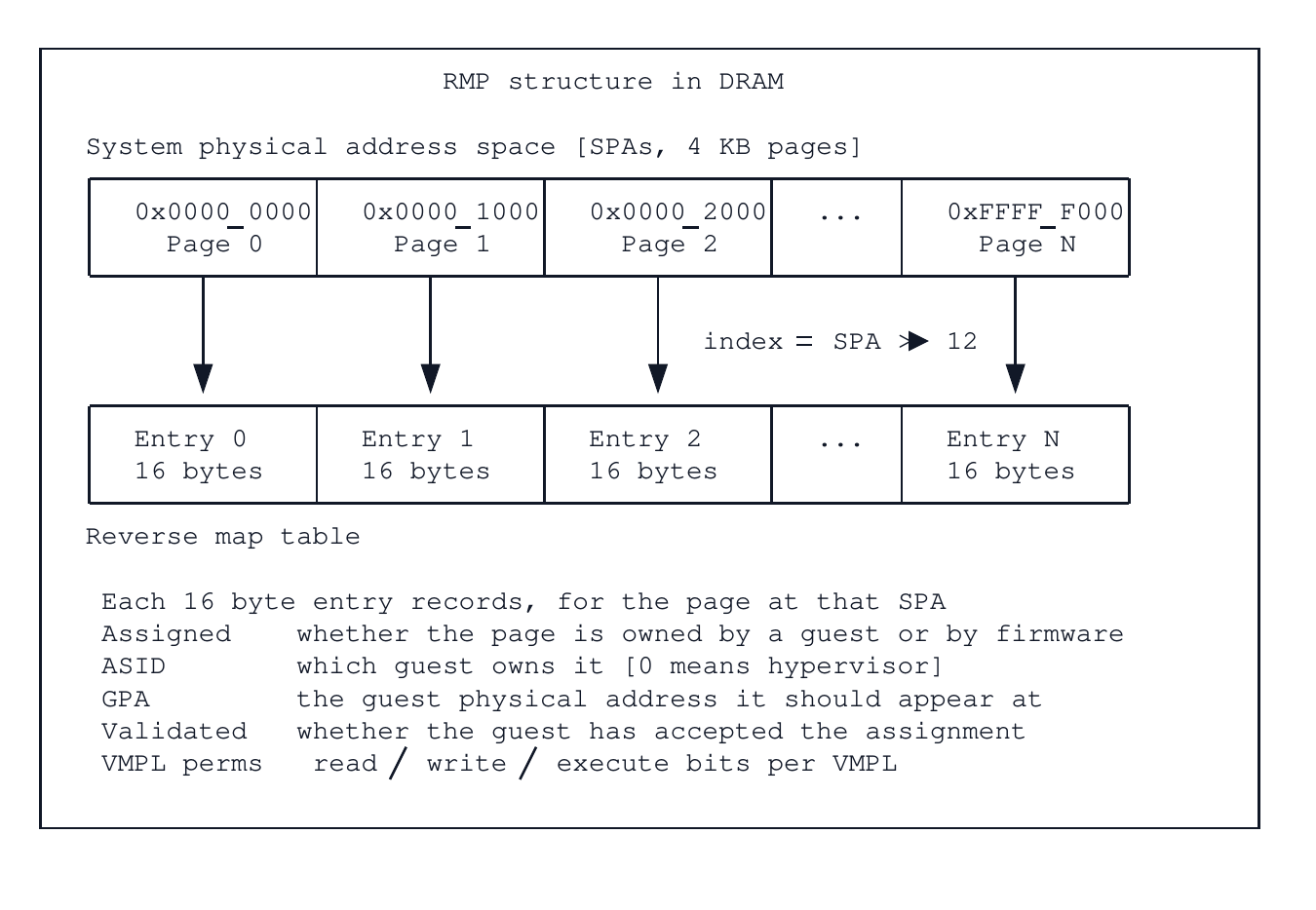}
\caption{The RMP in memory. The system physical address space is divided into 4 KB pages; the RMP is a parallel array of 16-byte entries, one per page, indexed by SPA shifted right by twelve. Each entry records the page's owner, the GPA at which it should appear, whether the guest has accepted it, and per-VMPL permission bits.}
\label{fig:07-rmp-structure}
\end{figure}

\begin{figure}[H]
\centering
\includegraphics[width=0.92\linewidth]{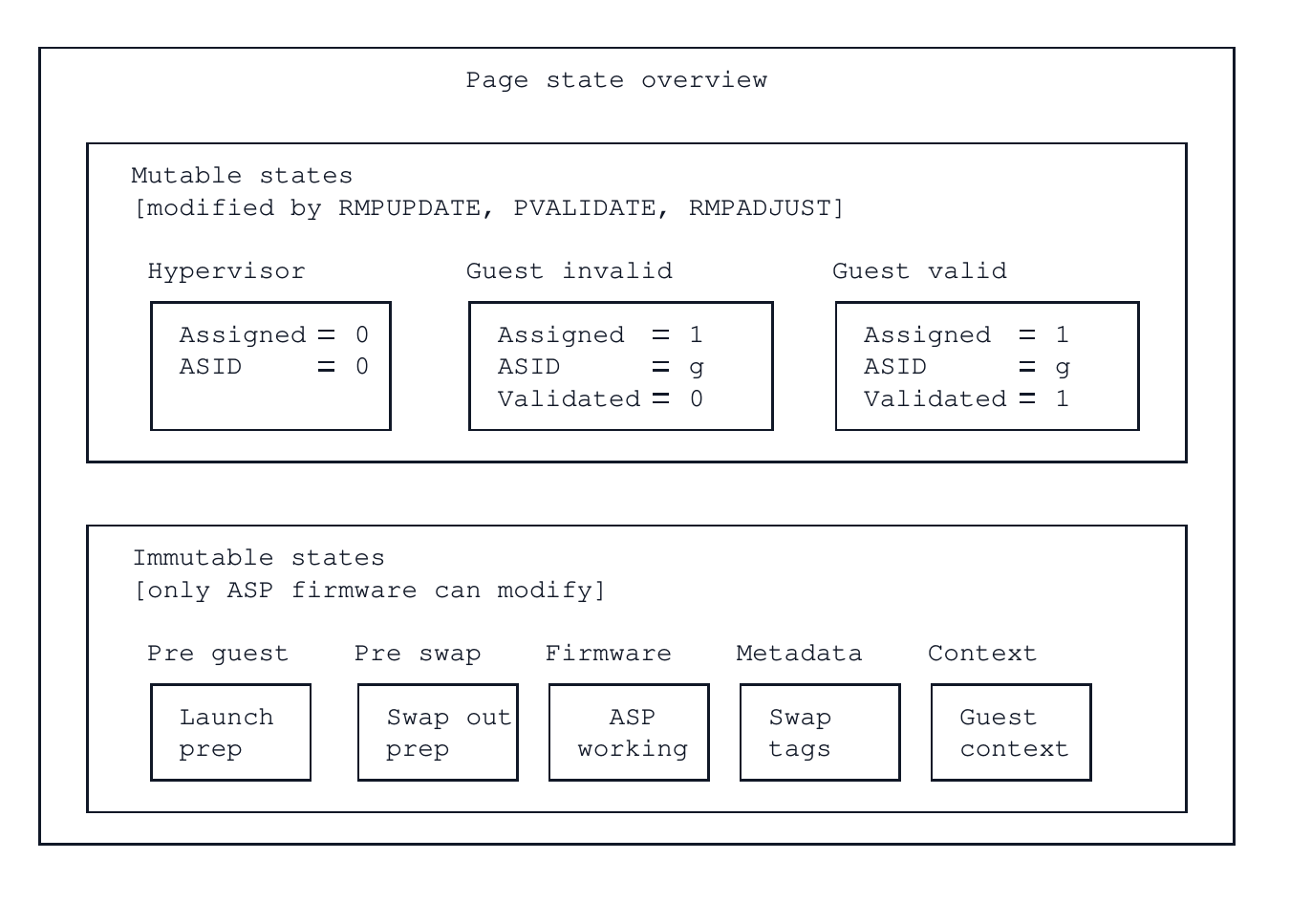}
\caption{RMP page states. The Hypervisor state is the default for unassigned memory. Guest invalid is the transitional state after a page has been assigned to a guest but before the guest has explicitly accepted it. Guest valid is the normal operating state for guest private memory. Immutable states are reserved for specific firmware operations (launch, swap, guest context) and can be modified only by the ASP.}
\label{fig:08-rmp-page-states}
\end{figure}

\subsection{The RMP Check}
\label{subsec:the-rmp-check}

Hardware performs an RMP check at the end of address translation for every relevant memory access. The check is inline with translation. The hardware walks the guest page tables, then the nested page tables, then consults the RMP entry for the resulting SPA, then either completes the access or raises a fault.

\begin{figure}[H]
\centering
\includegraphics[width=0.92\linewidth]{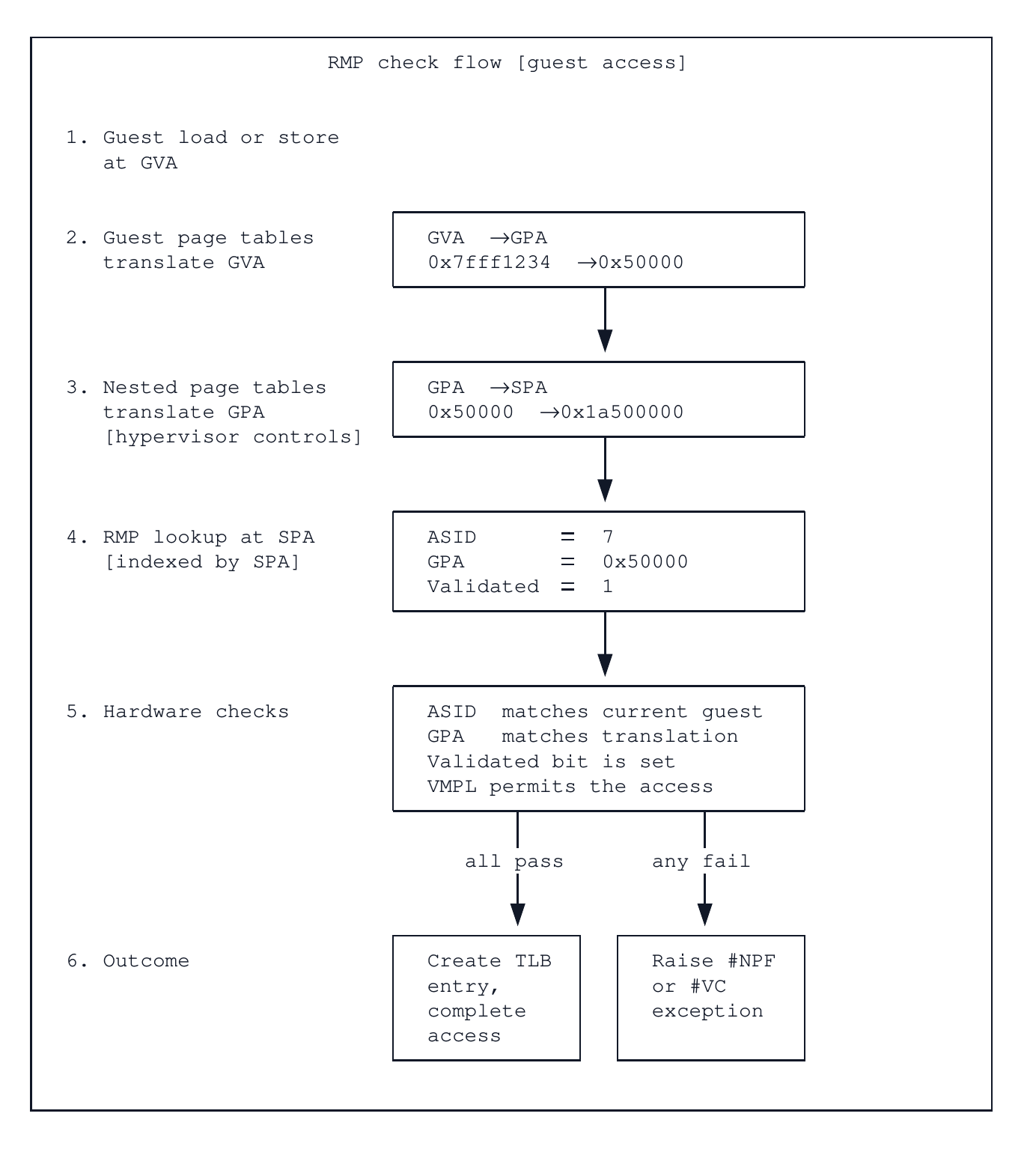}
\caption{The RMP check. Translation produces an SPA, the RMP entry for that SPA is consulted, and the hardware verifies that the ASID, GPA, validation bit, and VMPL permissions match the access being performed. The GPA comparison is the load-bearing check for remapping protection: the RMP records what GPA the page is supposed to appear at, and a translation that produces a different GPA fails the check.}
\label{fig:09-rmp-check-flow}
\end{figure}

When RMP checks occur depends on access type:

\begin{table}[H]
\centering
\begin{tabular}{lll}
\toprule
\textbf{Access type} & \textbf{RMP check?} & \textbf{Why} \\
\midrule
Guest private read [C=1] & Yes & Verifies ownership and validation \\
Guest private write [C=1] & Yes & Verifies ownership and validation \\
Guest shared access [C=0] & No & Shared pages are not RMP-protected \\
Hypervisor read of guest page & No & Encryption already protects content \\
Hypervisor write to guest page & Yes & Required to prevent corruption attacks \\
Page table A/D bit updates & Yes & Any write to a guest page is checked \\
\bottomrule
\end{tabular}
\end{table}

Shared pages bypass RMP checks because they are explicitly meant to be accessible to both guest and hypervisor. The guest chooses which pages to mark shared, typically for I/O and for the GHCB.

\subsection{Page Validation}
\label{subsec:page-validation}

The RMP tells hardware what the mapping should be. The question is how the RMP gets populated correctly in the first place. The answer is a two-step process that requires explicit cooperation from both hypervisor and guest.

\begin{figure}[H]
\centering
\includegraphics[width=0.92\linewidth]{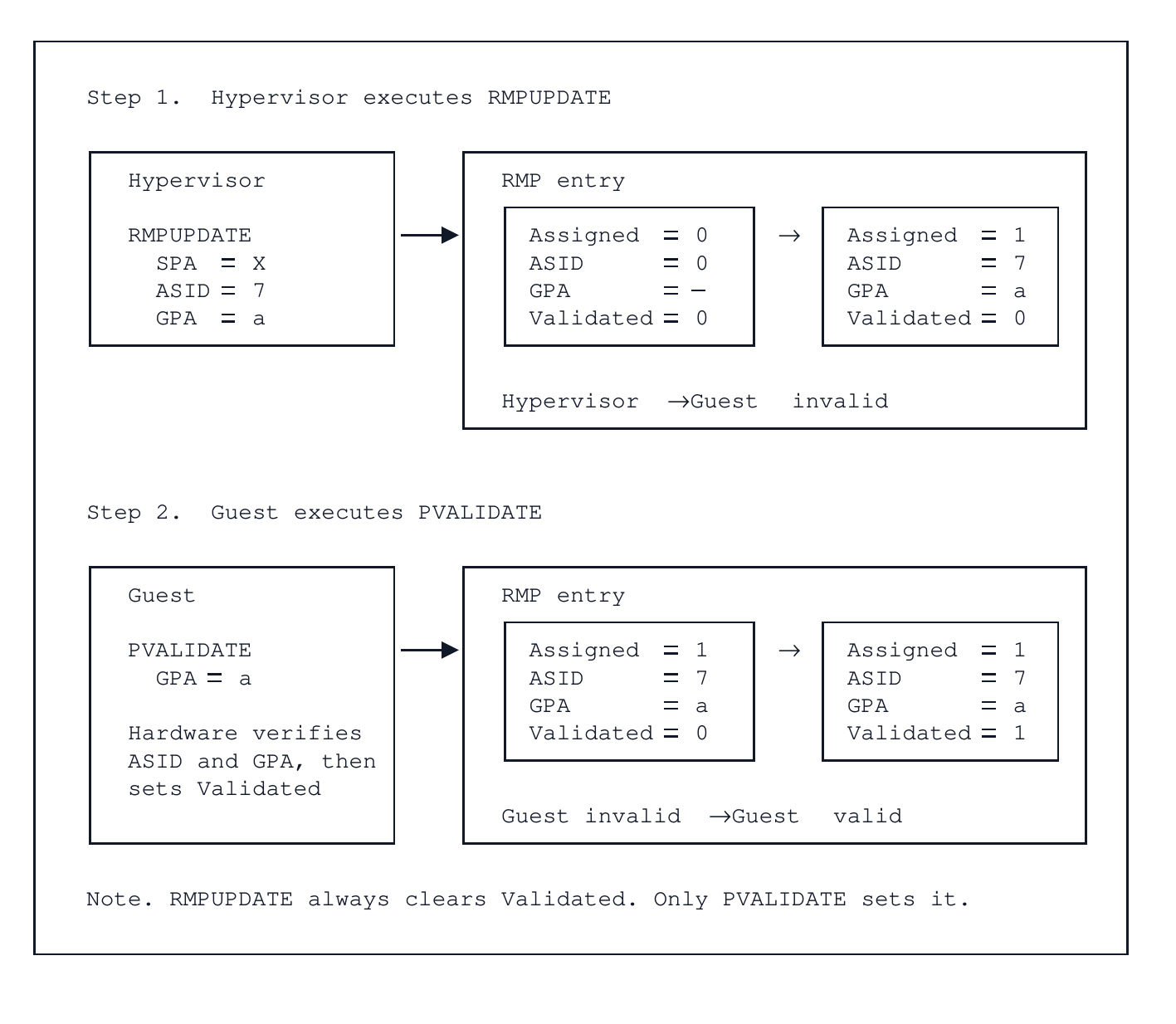}
\caption{The two-step page assignment. The hypervisor uses RMPUPDATE to assign a physical page to a specific guest at a specific GPA, leaving the page in the Guest invalid state. The guest then uses PVALIDATE to accept the assignment, transitioning the page to Guest valid. The hypervisor cannot validate pages on the guest's behalf.}
\label{fig:10-page-assignment}
\end{figure}

The note at the bottom of the figure is the load-bearing rule. RMPUPDATE always clears the Validated bit, regardless of the bit's previous value, and this is enforced in hardware. The reason is that if RMPUPDATE could preserve Validated, the hypervisor could wait for the guest to validate a page, then reassign that page to a different GPA while keeping Validated=1, and the guest would access the wrong physical page without knowing. By forcing Validated=0 on every RMPUPDATE, the hardware ensures that any change to page assignment requires the guest to re-validate, which puts the guest in a position to detect the change.

The corresponding rule on the guest side is equally strict: \textbf{the guest must never validate the same GPA twice.} If the guest validates GPA A, and later (after a Validated=0 fault) validates GPA A again, the integrity guarantee breaks. After a validation fault on a previously validated GPA, re-running PVALIDATE would accept whatever new SPA the hypervisor has installed, leaving the hypervisor free to switch the GPA between the original and substituted page on each access. The guest's correct response is to terminate or enter a safe state, not to re-validate.

Three instructions modify RMP entries.

\begin{itemize}
  \item \textbf{RMPUPDATE} (hypervisor instruction). Assigns pages to guests or reclaims them, sets ASID, GPA, and VMPL permissions, and always clears Validated. Cannot modify immutable pages.
  \item \textbf{PVALIDATE} (guest instruction, VMPL0 only). Validates or invalidates pages owned by the current guest. Hardware verifies that the RMP entry's ASID matches the current guest and that the entry's GPA matches the address being validated.
  \item \textbf{RMPADJUST} (guest instruction). Modifies the per-VMPL permissions for pages owned by the current guest, subject to the delegation rules in Section 5.1, so it can only lower the permissions granted to less privileged VMPLs and never raise its own. Any VMPL can execute it. Does not affect the Validated bit. The one form that is VMPL0 only is creating a VMSA page, which is why an SVSM must proxy vCPU creation for a guest OS at a lower VMPL.
\end{itemize}

When the guest runs an SVSM at VMPL0 (Section 5), the guest OS at VMPL2 cannot execute PVALIDATE directly and must call into the SVSM, and the same holds for the VMPL0-only operations RMPADJUST reaches, such as creating the VMSA pages for new vCPUs. This indirection is the price of the privilege separation, and it allows the SVSM to enforce additional policies, such as zeroing pages before handing them to the guest OS.

\subsection{Attack Detection Walkthrough}
\label{subsec:attack-detection-walkthrough}

The mechanism is most easily understood by tracing a specific attack. Consider a remapping attack. The guest writes a secret to GPA A, and the hypervisor wants the guest to later read different content from that same GPA. With the RMP in place, the attack proceeds as follows:

\begin{enumerate}
  \item \textbf{Initial state.} The NPT maps GPA A to SPA X. The RMP entry for X records ASID = guest, GPA = A, Validated = 1. The guest writes a secret at GPA A, and the ciphertext is stored at SPA X.
  \item \textbf{Attack setup.} The hypervisor runs RMPUPDATE with SPA = Y, ASID = guest, GPA = A, producing a fresh RMP entry for Y with Validated = 0. It then changes the NPT so that GPA A now translates to SPA Y. SPA X still has Validated = 1, but nothing points to it.
  \item \textbf{Guest access after the swap.} The guest reads GPA A. Translation produces SPA Y, and the RMP lookup finds the new entry. ASID and GPA both match, but Validated = 0, so the check fails and the hardware raises a \#VC exception.
  \item \textbf{Detection by the guest.} The guest sees a \#VC fault for a GPA it already validated. The only way that can happen is if the hypervisor changed the backing page underneath it. The correct response is to terminate.
\end{enumerate}

By the same chain of reasoning, replay is detected because the RMP records GPAs alongside ownership. Replaying old ciphertext requires either reusing the original SPA (in which case the hypervisor must use RMPUPDATE to reassign it, clearing Validated and prompting detection) or pointing the NPT at a different SPA (in which case the new SPA's RMP entry has the wrong GPA or Validated=0). Aliasing is detected because each RMP entry records exactly one expected GPA, so a second GPA mapped to the same SPA cannot match. Corruption is detected because the corrupted ciphertext decrypts to garbage rather than to plaintext that survives application-level integrity checks.

\subsection{The Bijective Mapping Property}
\label{subsec:the-bijective-mapping-property}

The RMP and validation discipline together establish a one-to-one correspondence between the guest's view of physical addresses and the system's. Concretely:

\begin{itemize}
  \item An SPA can be the target of at most one GPA in the active mapping. This is enforced by the RMP's GPA field: if SPA X has RMP[X].GPA = A, then a translation that maps a different GPA B to SPA X (with B not equal to A) fails the GPA check.
  \item A GPA can be backed by at most one SPA at any given time, in the sense that any change to which SPA backs a GPA requires RMPUPDATE on the new SPA, which clears Validated, which the guest detects.
\end{itemize}

\begin{figure}[H]
\centering
\includegraphics[width=0.92\linewidth]{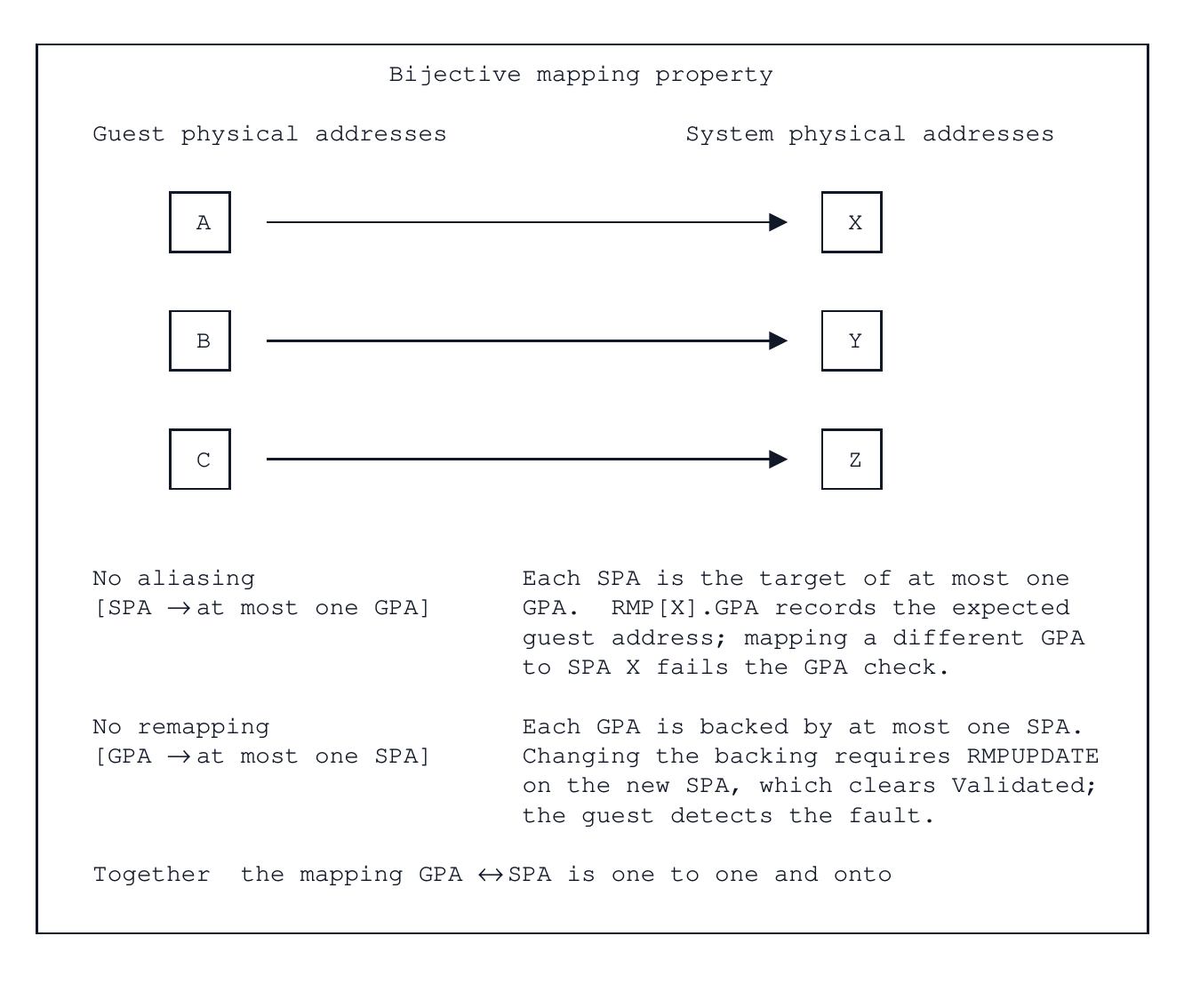}
\caption{The GPA-to-SPA mapping is bijective. The RMP's GPA field rules out aliasing (two GPAs cannot legitimately point at the same SPA) and the validation discipline rules out remapping (changing the SPA behind a GPA invalidates the new SPA). The composition is the integrity property the guest sees: reads return what was last written.}
\label{fig:13-bijective-mapping}
\end{figure}

The consequence is that the guest's memory abstraction holds. Reading an address returns what was last written there, never stale data, never another page's data. This is the integrity property SEV-SNP guarantees, expressed at the level of the guest's view rather than at the level of the underlying mechanism.

\subsection{IOMMU Integration}
\label{subsec:iommu-integration}

The RMP is not consulted only by CPUs. Devices performing DMA also pass through an RMP check, performed by the IOMMU. Without this, a confused-deputy attack would be available. The hypervisor would program a device to DMA to a guest-private SPA, and the device, executing outside any guest context, would read or write the encrypted page directly.

With IOMMU RMP checks in place, a device DMA attempt is blocked when the target SPA is guest-private (Assigned=1, ASID set). Device DMA is permitted to hypervisor pages (Assigned=0) and to shared guest pages (C=0). This is why guests use bounce buffers for I/O: data moves through shared pages that devices can access, then is copied to and from private pages by the guest.

Current parts relax this for devices that support Trusted I/O (TIO), AMD's SEV extension built on the PCI-SIG TDISP standard \cite{ref21}. Under TIO, a device authenticates itself to the guest (using SPDM, with the PCIe link protected by IDE), the guest verifies the device's identity and firmware, and the device is bound into the guest's trust domain; the IOMMU then permits that device to DMA directly into guest-private (C=1) memory. The data path no longer transits shared pages, eliminating both the bounce-buffer copies and the need for application-level encryption on that path, and the device's trustworthiness is established by attestation rather than by trusting the hypervisor's mediation. TIO requires support in the platform, the device, and the guest, and its presence is reported in the attestation report (Section 6.2); deployments without it continue to use bounce buffering as described above.

\subsection{What RMP Protection Does Not Cover}
\label{subsec:what-rmp-protection-does-not-cover}

The RMP provides strong integrity guarantees within a precisely scoped boundary, and it is worth being explicit about what falls outside that boundary.

\begin{itemize}
  \item \textbf{Denial of service.} The hypervisor can refuse to schedule the guest, refuse to provide memory, or provide memory that is intentionally slow to access. Availability is out of scope.
  \item \textbf{Side channels.} The hypervisor can observe which pages the guest accesses by watching access and dirty bits in page tables, by monitoring nested page faults, or by analyzing cache and timing behavior. This leaks information about access patterns, though not the underlying data. On parts without ciphertext hiding, or where it is disabled, the ciphertext side channel also remains open (Section 3.3); enabling ciphertext hiding closes that particular channel at the memory controller.
  \item \textbf{Shared page manipulation.} Shared (C=0) pages bypass the RMP. The guest must not place sensitive data in shared pages and must validate any data received through them.
  \item \textbf{Bugs in the guest.} A vulnerability in guest software is exploitable in the usual way. The RMP protects the environment, not the code running in it.
\end{itemize}

These exclusions are consistent with the threat model in Section 2. The boundary is sharp and meaningful, but it is not a defense against everything an adversary might attempt.

\section{Privilege and Communication}
\label{sec:privilege-and-communication}

The mechanisms in Sections 3 and 4 produce a guest whose memory and registers are inaccessible to the hypervisor and whose memory cannot be silently tampered with. They do not, by themselves, allow the guest to do useful work. A guest still needs the hypervisor for things only the hypervisor can provide: device I/O, certain CPU services, memory provisioning, and information about the platform on which it is running. In conventional virtualization, these requests are serviced by the hypervisor reading guest state directly. Under SEV-SNP, that state is encrypted, so the cooperation protocol has to be redesigned.

This section covers the three mechanisms that make the redesigned protocol work. VM Privilege Levels provide privilege separation inside an encrypted guest. The encrypted VM Save Area protects CPU state across exits. The GHCB protocol is the explicit channel through which a guest cooperates with an untrusted hypervisor. The section also covers the optional interrupt-protection modes that close a residual exposure on the interrupt-injection path.

\subsection{VM Privilege Levels}
\label{subsec:vm-privilege-levels}

VMPLs solve a different problem from the guest-hypervisor cooperation question. They address what privilege separation should look like inside the encrypted guest itself.

Without VMPLs, every component inside an SNP guest runs at the same hardware privilege from the SEV-SNP perspective. The guest kernel can access all guest memory; any code in the guest can execute any instruction the guest is allowed to execute. This is fine when everything in the guest is trusted equally. It is not fine when the deployment includes a security monitor that even a compromised guest kernel should not be able to tamper with, a virtual TPM whose keys must be inaccessible to the guest OS, or an unmodified legacy OS that does not understand SEV-SNP and needs a small trusted shim to handle the SNP-specific operations on its behalf.

VMPLs are a hardware-enforced privilege hierarchy with four levels (0 through 3, with VMPL0 the highest privilege). Each virtual CPU runs at exactly one VMPL at any given time, and the VMPL determines which memory the vCPU can access and which instructions it can execute. VMPLs are an additional privilege axis, separate from the conventional ring 0 / ring 3 distinction; a process can be in ring 0 (kernel mode) at VMPL2, or in ring 3 (user mode) at VMPL0, in any combination.

VMPLs are enforced through the same RMP that tracks page ownership. Each RMP entry has four sets of permission bits, one per VMPL, granting some combination of read, write, and execute. When code at VMPL2 accesses a page, the hardware checks the VMPL2 permission bits in that page's RMP entry; if the bits do not allow the access, the CPU raises an exception.

\begin{figure}[H]
\centering
\includegraphics[width=0.7\linewidth]{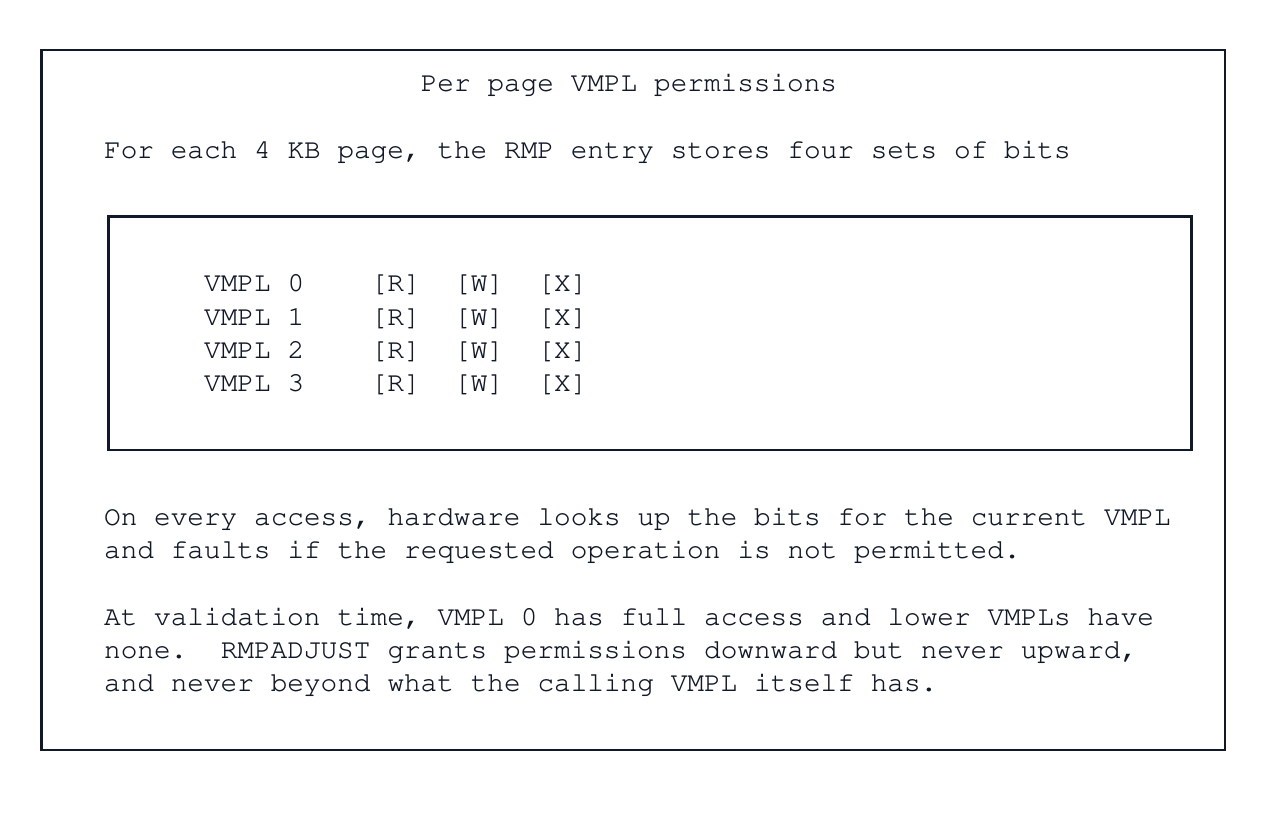}
\caption{Per-page VMPL permissions. Each RMP entry stores four (read, write, execute) triples, one per VMPL. On every guest access, hardware looks up the triple for the current VMPL and faults if the requested operation is not permitted. RMPADJUST grants permissions downward only and never beyond what the calling VMPL itself holds.}
\label{fig:14-vmpl-perms}
\end{figure}

When a page is first validated, VMPL0 has full access and lower VMPLs have none. VMPL0 must explicitly grant permissions to lower levels using RMPADJUST. RMPADJUST is restricted in three ways: it can only grant permissions to less-privileged VMPLs (VMPL0 can grant to VMPL1/2/3, VMPL2 can grant only to VMPL3), it cannot grant more permission than the current VMPL has, and it cannot increase the current VMPL's own permissions. The result is a strict top-down delegation model.

\subsubsection{The SVSM}

The natural question is what runs at VMPL0. In most non-trivial deployments, the answer is the SVSM (Secure VM Service Module) \cite{ref22}, a small piece of trusted code that runs at the highest VMPL inside the guest, is measured during launch, and provides services to the guest OS running at a lower VMPL.

\begin{figure}[H]
\centering
\includegraphics[width=0.92\linewidth]{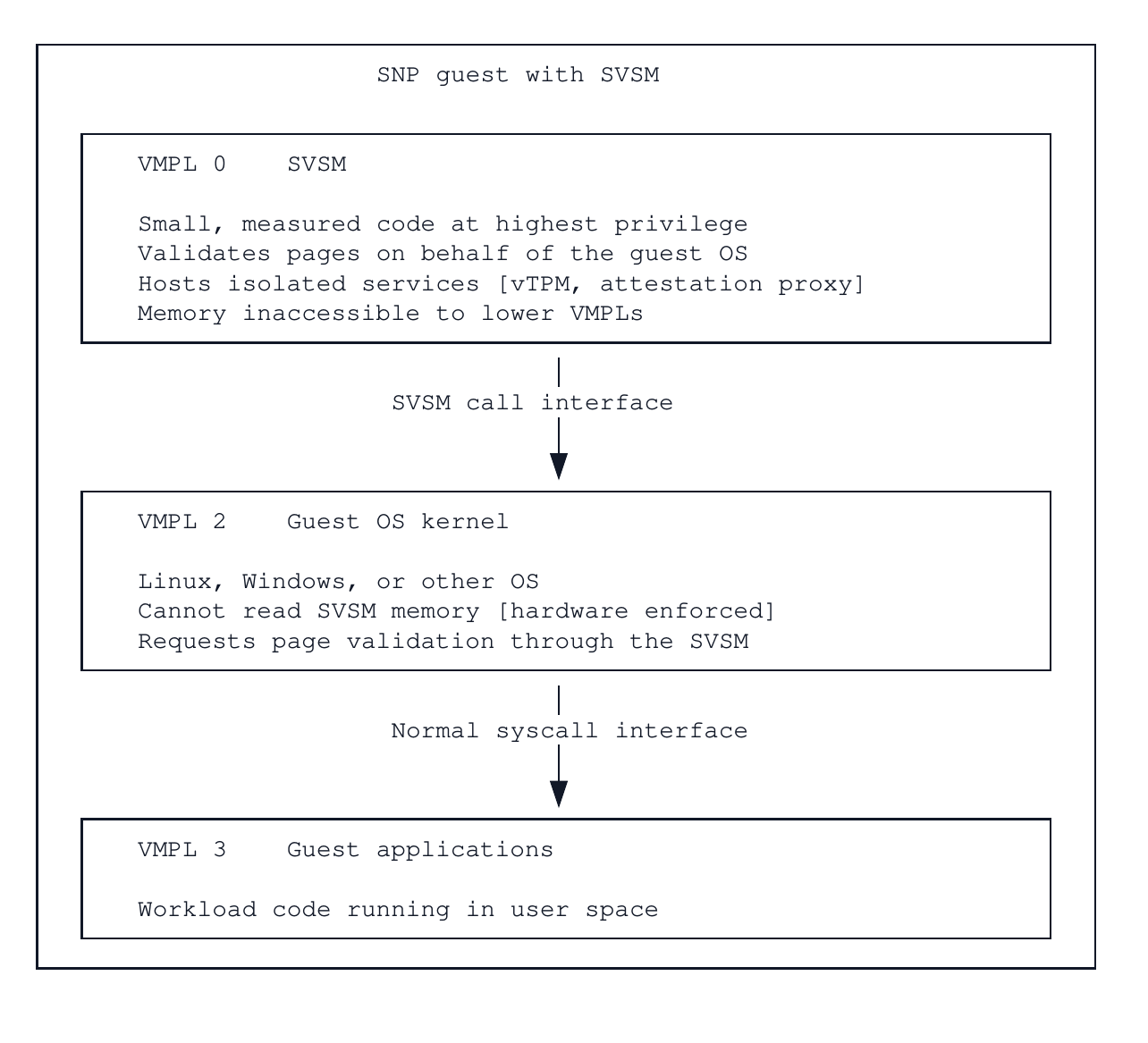}
\caption{An SNP guest with an SVSM. The SVSM occupies VMPL0 and proxies VMPL0-only operations on behalf of a guest OS running at VMPL2. The guest applications run at VMPL3 in the normal way. The SVSM's memory has VMPL0-only permissions, so neither the guest OS nor applications can read or write it, even with kernel-level privilege.}
\label{fig:15-vmpl-svsm-stack}
\end{figure}

The SVSM is necessary because certain operations can only be performed at VMPL0.

\begin{itemize}
  \item \textbf{PVALIDATE.} Page validation (Section 4) requires VMPL0. A guest OS at VMPL2 cannot validate pages directly and must call into the SVSM.
  \item \textbf{Creating new vCPU contexts.} Bringing additional virtual CPUs online means creating their VMSA pages, and creating a VMSA page with RMPADJUST is a VMPL0-only operation. RMPADJUST itself is not VMPL0 only. It is the general downward-delegation instruction of Section 5.1, and only this VMSA-creating form is restricted to VMPL0.
\end{itemize}

The SVSM proxies these operations. The guest OS makes a request through a defined calling convention; the SVSM verifies the request against its own policy (for example, that the requested page is one the guest OS is allowed to validate) and then performs the operation. The SVSM's memory has VMPL0-only permissions, so a compromised guest kernel cannot read SVSM code, modify it, or extract any secrets it holds.

Not all SEV-SNP deployments use an SVSM. In the simpler \emph{enlightened guest} model, the guest OS itself runs at VMPL0 and handles SNP operations directly. This is operationally simpler but loses the privilege separation, since a kernel exploit gives the attacker the same privilege as the SNP-aware code. Deployments that need a vTPM, that want defense in depth against kernel exploits, or that want to run an unmodified legacy OS with a small trusted shim use the SVSM model.

\subsection{Protecting CPU State: the VMSA}
\label{subsec:protecting-cpu-state-the-vmsa}

When a virtual machine pauses (because the hypervisor needs to run, or because an exception occurred, or because an external interrupt arrived), the CPU's current state has to be saved somewhere. This saved state is the VMSA (VM Save Area). It carries the entire execution context of the vCPU at the moment of the exit.

The VMSA holds, at minimum: the instruction pointer (where execution will resume), the stack pointer (where the current stack lives), the general-purpose registers (which often hold sensitive data, function arguments, intermediate values, or in-flight cryptographic material), the segment registers, and the control registers (CR3 in particular controls the page table base; modifying it would remap memory).

In conventional virtualization, the hypervisor reads and writes all of this. SEV-ES and SEV-SNP make the VMSA opaque. Three protections apply.

First, the VMSA is encrypted with the guest's VEK. The hypervisor sees ciphertext when it reads the VMSA page.

Second, the RMP marks VMSA pages specially (the VMSA bit in the RMP entry), and the hypervisor cannot write to them. Hardware integrity-protects the VMSA in the same way it protects guest private memory.

Third, only specific hardware operations (the hardware's own save and restore on VM entry and exit) can modify the VMSA. Software, including the guest itself at lower VMPLs, cannot directly write it.

\begin{figure}[H]
\centering
\includegraphics[width=0.92\linewidth]{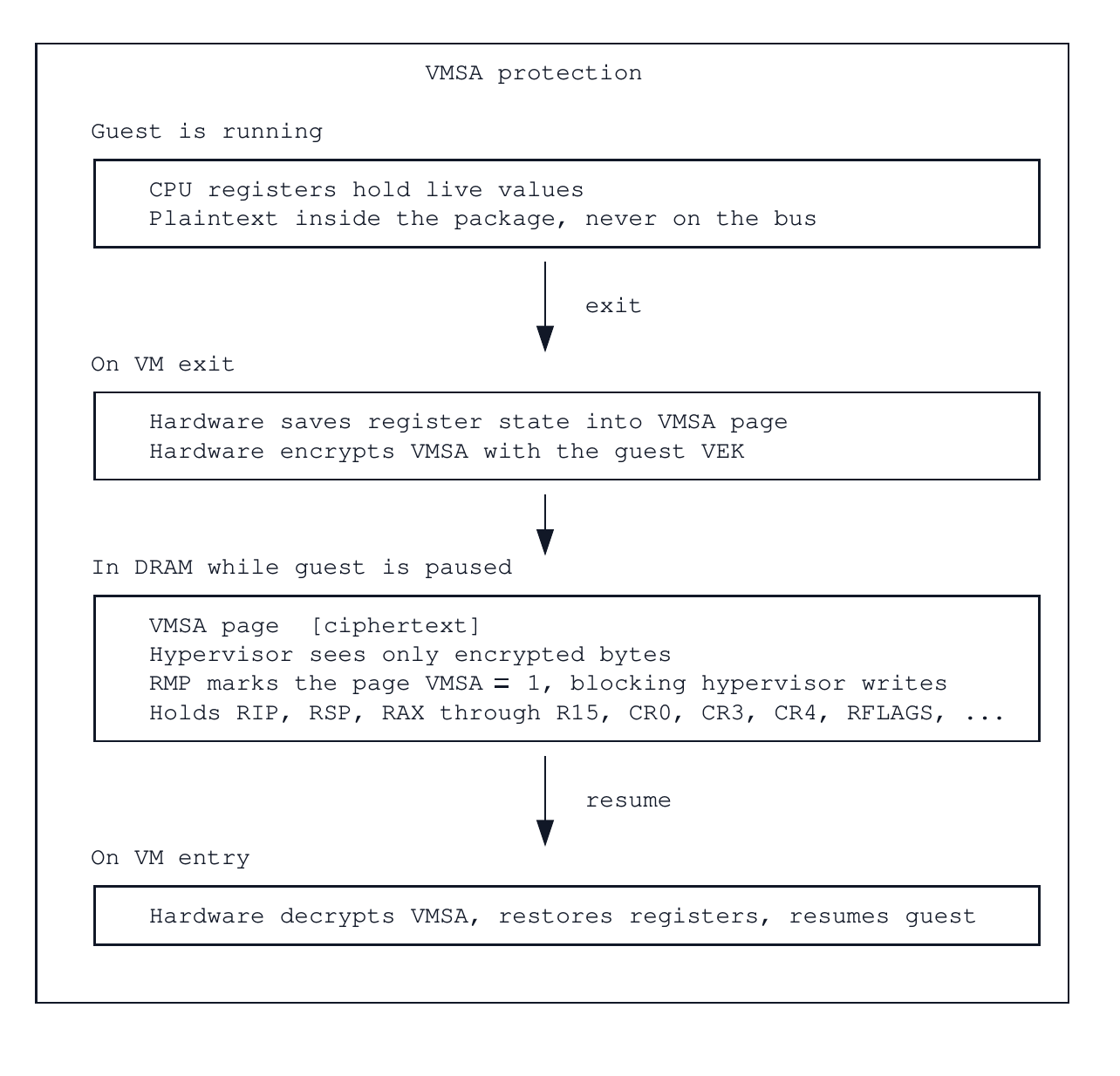}
\caption{How the VMSA is protected across an exit and re-entry. While the guest runs, registers hold plaintext values inside the package. On exit, hardware saves the registers into the VMSA page and encrypts the page with the guest VEK; the RMP marks the page as VMSA so the hypervisor cannot write it. On re-entry, hardware decrypts the VMSA and restores registers, with no software path through the hypervisor.}
\label{fig:16-vmsa-protection}
\end{figure}

When an SNP guest runs with VMPL separation, there can be more than one VMSA per vCPU: a VMSA for VMPL0 (the SVSM) and a separate VMSA for VMPL2 (the guest OS) on the same vCPU. This allows the hardware to switch between privilege levels while keeping each level's state separate.

The hypervisor still has the VMCB control area to work with, so it reads exit codes, configures intercepts, and learns what kind of exit happened, but the detailed register state that classical virtualization would have made visible is no longer available.

\subsection{Guest-Hypervisor Communication}
\label{subsec:guest-hypervisor-communication}

With the VMSA encrypted, a new question arises. The guest still needs the hypervisor to handle CPUID, MSR access, port I/O, and other operations that conventional virtualization handled by the hypervisor inspecting guest registers. How does an encrypted guest communicate with an untrusted hypervisor when the hypervisor cannot see what the guest is asking?

The answer is the \#VC exception and the GHCB protocol. The design is structured around two principles. First, the guest controls what the hypervisor sees. Instead of the hypervisor automatically observing all guest state, the guest explicitly shares only what is needed for each operation. Second, the guest validates everything the hypervisor returns. A malicious hypervisor might lie, and the guest checks responses against known-good values before trusting them.

\subsubsection{The \#VC exception}

When the guest executes an instruction that previously would have caused an exit visible to the hypervisor, the hardware raises a \#VC (VMM Communication) exception inside the guest. The exception is exception vector 29, introduced with SEV-ES and inherited by SEV-SNP. The guest's \#VC handler runs in the guest, decides what to share with the hypervisor, prepares a message in a shared page (the GHCB), and only then performs an explicit exit (VMGEXIT) to ask the hypervisor for the operation.

The flow is:

\begin{enumerate}
  \item The guest executes an intercepted instruction (CPUID, RDMSR, port I/O, ...).
  \item Hardware raises \#VC inside the guest. There is no automatic exit to the hypervisor.
  \item The guest's \#VC handler runs. It inspects the offending instruction, decides what state to share, and writes a request into the GHCB shared page.
  \item The guest issues VMGEXIT, a voluntary exit.
  \item The hypervisor reads the GHCB, performs the operation, and writes the response back.
  \item The guest resumes. The \#VC handler validates the response: CPUID against the measured CPUID page, MSRs against architectural ranges, and so on.
  \item The handler returns a validated result to the original code.
\end{enumerate}

The guest controls the information flow. The hypervisor sees only what the \#VC handler has placed in the GHCB, and only after a voluntary VMGEXIT.

\subsubsection{The GHCB}

The GHCB (Guest-Hypervisor Communication Block) is a 4 KB page marked shared (C=0) so that both the guest and the hypervisor can read and write it. Its layout is defined by the GHCB specification \cite{ref23}. It includes an exit code (what operation the guest is requesting), exit information (additional parameters), space for register values that the operation needs to communicate, and a valid bitmap that records which fields the guest has actually filled in. The bitmap matters because the guest marks only the fields it needs to share, minimizing information exposure.

\begin{figure}[H]
\centering
\includegraphics[width=0.92\linewidth]{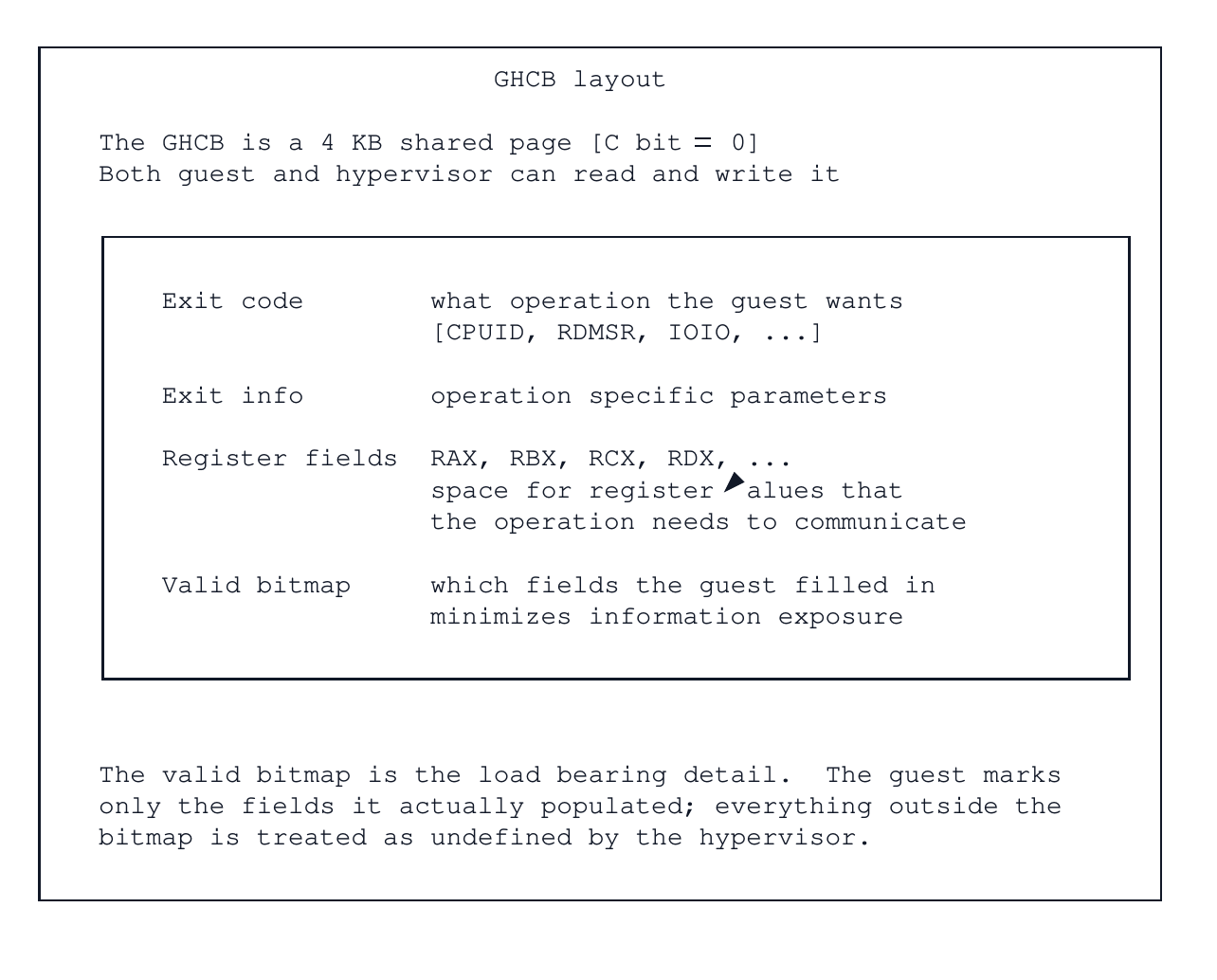}
\caption{GHCB layout. The shared page carries an exit code identifying the requested operation, operation-specific parameters, register fields for values the operation needs, and a valid bitmap that names which register fields the guest has filled in. Anything outside the valid bitmap is undefined; the bitmap is the mechanism by which the guest minimizes information exposure on each request.}
\label{fig:18-ghcb-layout}
\end{figure}

\begin{figure}[H]
\centering
\includegraphics[width=0.98\linewidth]{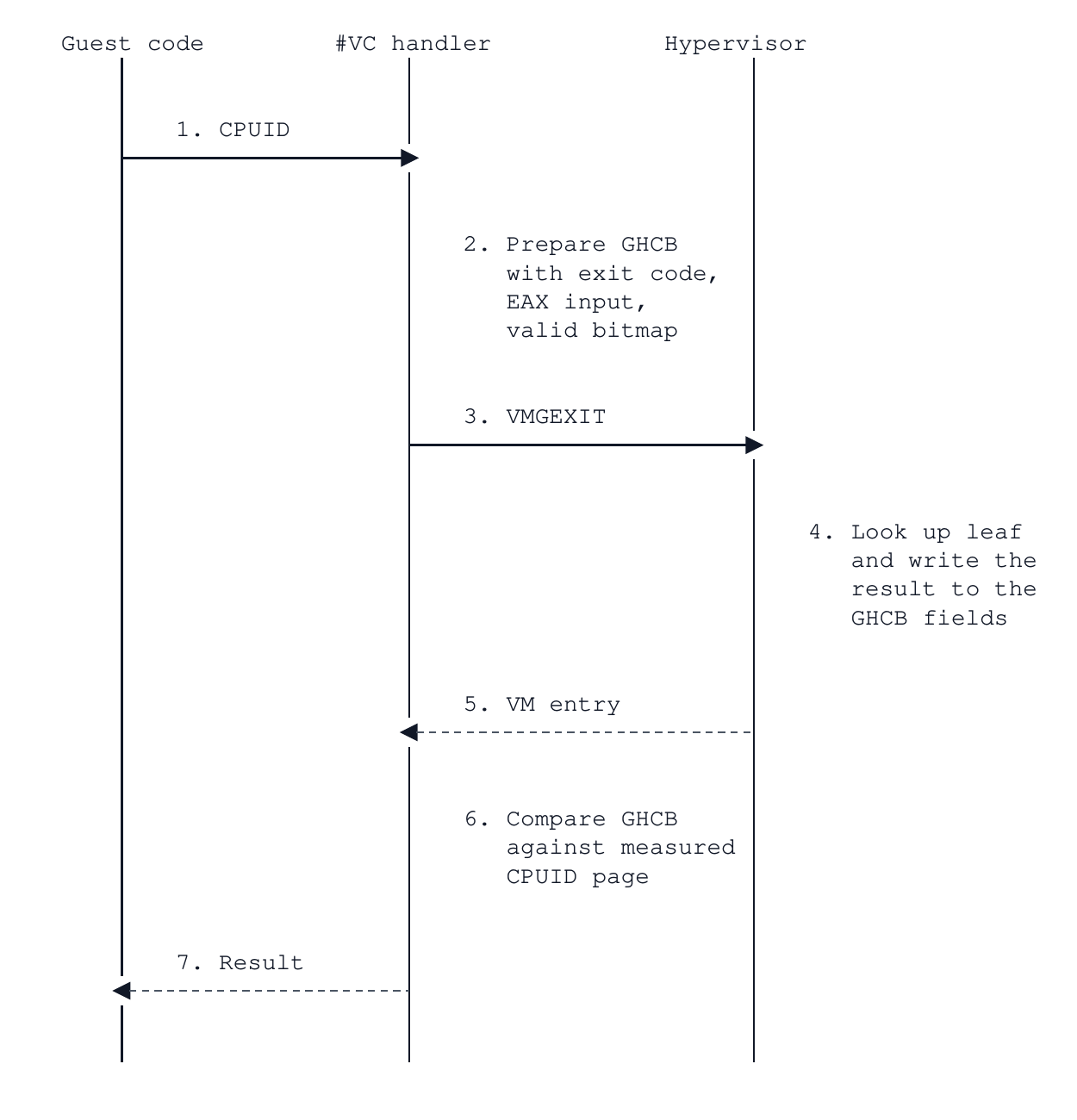}
\caption{A CPUID request handled through the \#VC and GHCB mechanism. The guest places the request in the GHCB and exits to the hypervisor. The hypervisor reads the request, writes a response, and returns. The guest validates the response against a measured CPUID page before believing it. Step 6 is the load-bearing one. Without validation the hypervisor could lie about CPU features in ways the guest would silently accept.}
\label{fig:19-cpuid-vc-flow}
\end{figure}

The complete cycle for an example operation is shown in Figure~\ref{fig:19-cpuid-vc-flow}.

\subsubsection{Validating responses}

Different operations have different validation strategies.

For \textbf{CPUID}, the SNP launch process places a CPUID page into the guest at a known GPA. The CPUID page records the values the guest expects to see for every CPUID leaf the workload cares about. The page is part of the guest's launch image and therefore part of the launch measurement (Section 6). On every CPUID \#VC, the guest's handler compares the hypervisor's response against the measured page; a mismatch indicates either a bug or an attack and is treated as a fatal condition.

For \textbf{MSR reads}, validation is per-MSR: some MSRs have architecturally defined value ranges, some have known correct values for the platform, and some are inherently hypervisor-controlled and must be treated with suspicion.

For \textbf{I/O operations}, direct validation is generally not possible, because device responses are whatever the (hypervisor-mediated) device says. For sensitive data, the guest uses application-level encryption (TLS for network traffic, LUKS for disk) so that the device never sees plaintext.

For \textbf{memory allocation}, the RMP and page-validation mechanism (Section 4) handle the work. A page is not usable until the guest has explicitly accepted it, and re-assignment by the hypervisor is detected as a Validated=0 fault.

A malicious hypervisor still has options that the validation does not address. It can refuse to respond, hanging the guest indefinitely (denial of service, out of scope). It can return stale data by replaying old responses; this is mitigated by sequence numbers where applicable. It can modify the GHCB during processing; this is mitigated by the guest copying values out of the GHCB to private memory before relying on them. The guest must be paranoid about everything in the GHCB.

\subsection{Interrupt Protection}
\label{subsec:interrupt-protection}

The remaining communication channel that needs protection is interrupts. A conventional hypervisor can inject any interrupt into a guest at any time. A malicious hypervisor could inject interrupts when they are supposed to be disabled, inject spurious exceptions to confuse guest state machines, or manipulate interrupt priority to alter the guest's control flow. SEV-SNP provides two optional modes that close this channel, and current parts add a third, hardware-enforced mechanism.

In \textbf{restricted injection} mode, the hypervisor can only inject one specific exception, the \#HV (hypervisor exception). \#HV acts as a doorbell, telling the guest that something happened without specifying what. The guest's \#HV handler then consults a guest-managed event queue to determine what to do, and decides when and how to deliver each signal. The hypervisor is reduced from being able to inject arbitrary interrupts to being able only to ring a bell.

In \textbf{alternate injection} mode, the standard interrupt interface is preserved, but the control fields that determine what is injected live in the encrypted VMSA. Only code with VMSA write access (which is to say, VMPL0) can inject interrupts into lower VMPLs. The hypervisor still cannot inject directly.

The third mechanism, available on current parts, is Secure AVIC \cite{ref1,ref23}. The guest owns its per-vCPU interrupt-controller backing page and declares in a guest-controlled bitmap which interrupt vectors the hypervisor is permitted to deliver; hardware delivers only the intersection of what the hypervisor requests and what the guest allows, and NMIs additionally require a guest-armed request flag. A vector the guest never authorized is never delivered. This closes by construction the channel that restricted and alternate injection close by protocol, and it was motivated in part by attacks that abused the \#VC vector itself to confuse guest exception handlers \cite{ref24}. Secure AVIC composes with the modes above and, like them, does not constrain interrupt timing.

These modes compose with the SVSM. With both, the hypervisor signals VMPL0 via \#HV, the SVSM decides what to inject into the guest OS at VMPL2, and the guest OS sees what looks like a normal interrupt without needing to know about the indirection. The hypervisor's ability to manipulate interrupt timing remains; the hardware cannot prevent the hypervisor from delaying a doorbell. But it can no longer choose what the guest sees.

\subsection{The SVSM Calling Convention}
\label{subsec:the-svsm-calling-convention}

When the guest OS needs a service that only VMPL0 can provide (page validation, RMPADJUST, vTPM operations), it invokes the SVSM through a defined calling convention. The flow has to be designed against an adversary who controls scheduling. The hypervisor decides when each VMPL runs, and a malicious hypervisor could attempt to make it look as though the SVSM ran when it did not.

\begin{figure}[H]
\centering
\includegraphics[width=0.98\linewidth]{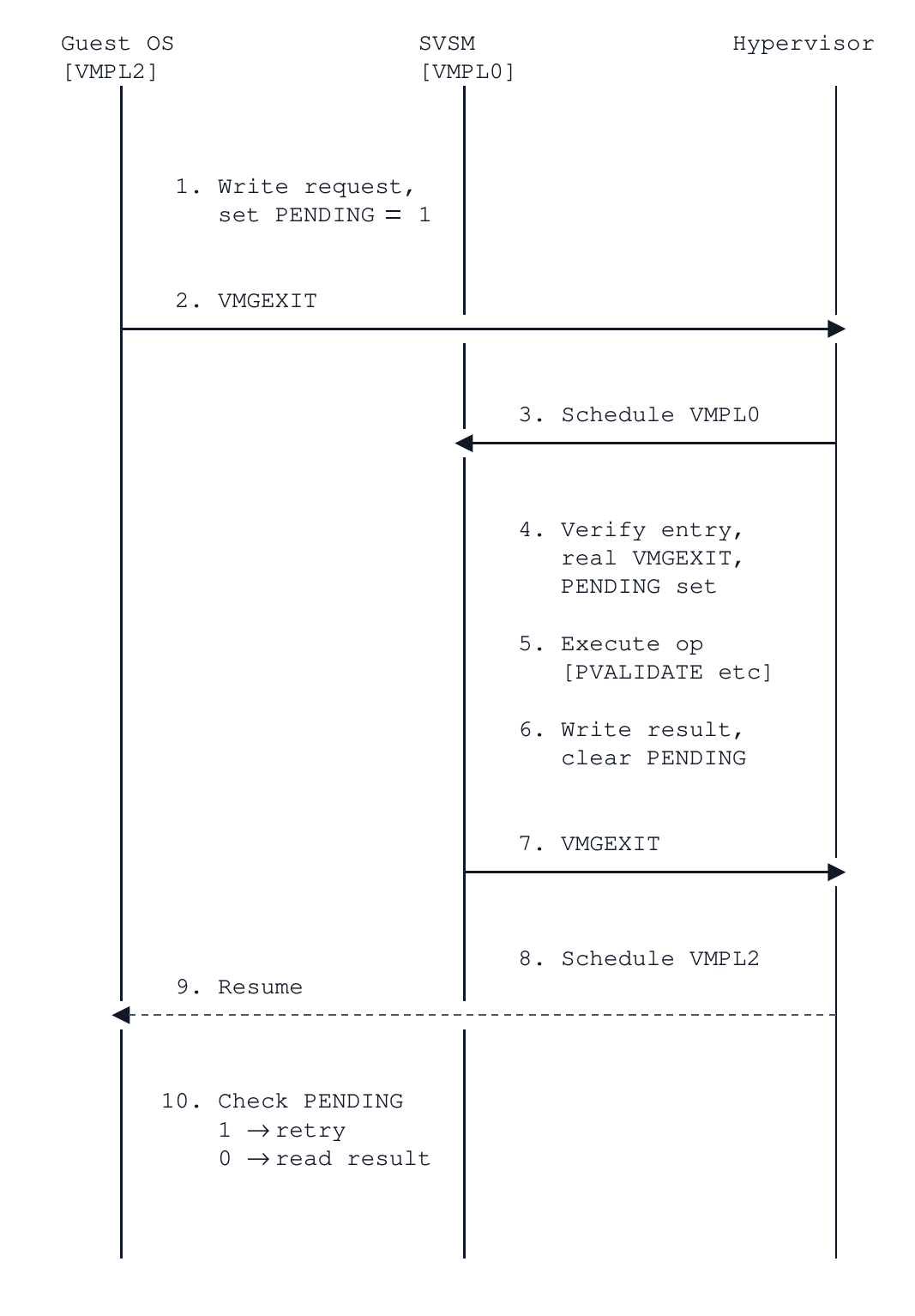}
\caption{SVSM call flow. The guest OS writes its request, sets a pending flag, and exits to the hypervisor. The hypervisor schedules the SVSM at VMPL0; the SVSM verifies that a real call is pending, executes the operation, writes the result, and clears the pending flag. After resuming, the guest OS checks whether the pending flag is still set: if so, the SVSM never actually ran and the call should be retried.}
\label{fig:21-svsm-call-flow}
\end{figure}

The pending flag is the load-bearing detail. The hypervisor controls scheduling and could return to the guest OS without ever running the SVSM. The guest OS detects this case by inspecting the pending flag after resuming. If the flag is still set, the SVSM was never given an opportunity to clear it, and the call should be retried.

The SVSM also zeros VMPCK0 (the VMPL0 communication key) from the secrets page after startup. This prevents the guest OS from impersonating VMPL0 when communicating with the ASP, since ASP messages from VMPL0 require VMPCK0. The guest OS uses VMPCK2 or VMPCK3 for its own ASP communication. The implication for attestation (Section 6) is that VMPL0 attestation reports can only be requested by the SVSM; the guest OS sees reports tagged with its own VMPL.

\begin{figure}[H]
\centering
\includegraphics[width=0.92\linewidth]{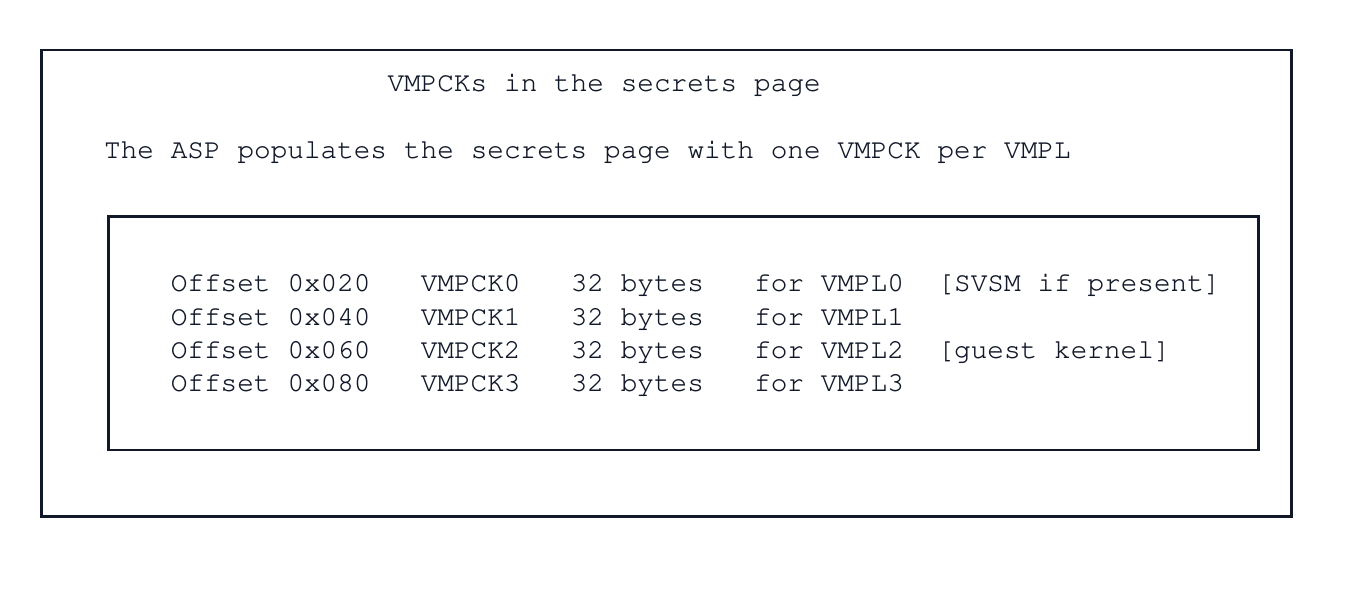}
\caption{VMPCKs in the secrets page. The ASP populates one AES-256 key per VMPL into the encrypted, RMP-immutable secrets page during launch. The hypervisor never sees these keys. Guest-to-ASP messages over the GHCB are authenticated and encrypted with AES-256-GCM under the appropriate VMPCK, and sequence numbers prevent replay.}
\label{fig:22-vmpck-secrets}
\end{figure}

The mechanisms in this section together describe how a SEV-SNP guest cooperates with an untrusted hypervisor without exposing state that the threat model wants to keep private.

\begin{itemize}
  \item \textbf{VMPLs} provide hardware-enforced privilege separation inside the encrypted guest, allowing trusted services such as an SVSM-hosted vTPM to run at higher privilege than the guest OS.
  \item \textbf{The encrypted VMSA} prevents the hypervisor from reading or modifying register state across exits.
  \item \textbf{The GHCB protocol} gives the guest explicit control over what information is shared with the hypervisor, and the guest validates responses where validation is possible (CPUID against a measured page, MSRs against architectural ranges).
  \item \textbf{Restricted and alternate injection} prevent the hypervisor from injecting arbitrary interrupts. The most it can do is ring a doorbell.
  \item \textbf{The SVSM calling convention} detects the case where the hypervisor returns control without actually running the SVSM, by requiring a pending flag that only the SVSM can clear.
\end{itemize}

What remains visible to the hypervisor under this design is timing (the hypervisor knows when exits occur and how long operations take), the contents of explicitly shared pages (the GHCB and any other C=0 pages), and the data path through I/O devices (which is why disk and network traffic carrying sensitive content is encrypted at the application level rather than relying on the TEE alone).

\section{Attestation}
\label{sec:attestation}

The mechanisms in Sections 3 through 5 constrain what an adversary can do to a running guest. They do not, on their own, prove anything to a remote party. A customer who deploys a workload on someone else's hardware needs cryptographic evidence that the workload is in fact running on the expected hardware, with the expected code, in a confidential environment. Attestation supplies that evidence.

Without attestation, a hypothetical attacker could run the workload unencrypted while merely claiming otherwise, modify the code before loading it, or run a completely different program that pretends to be the legitimate one. The verifier has no way to distinguish these cases from a correct deployment. Attestation closes this gap by binding a hardware-signed assertion (signed by a key rooted in the silicon itself) to the measurement of what was loaded into the guest and to data the guest chooses to commit to at report-generation time.

This section describes what is measured at launch, what an attestation report contains, the AMD key hierarchy under which reports are signed, the steps a verifier performs, and the optional Identity Block mechanism that lets a guest owner pre-commit to an expected measurement.

\subsection{The Launch Measurement}
\label{subsec:the-launch-measurement}

The launch measurement (the launch digest) is computed by the ASP as the guest is constructed. It captures the initial state of the VM at launch time: the bytes loaded into each guest physical address, the type of each page, the permissions assigned to each VMPL, and the order in which pages are installed. Three ASP commands cover the launch flow.

\begin{figure}[H]
\centering
\includegraphics[width=0.92\linewidth]{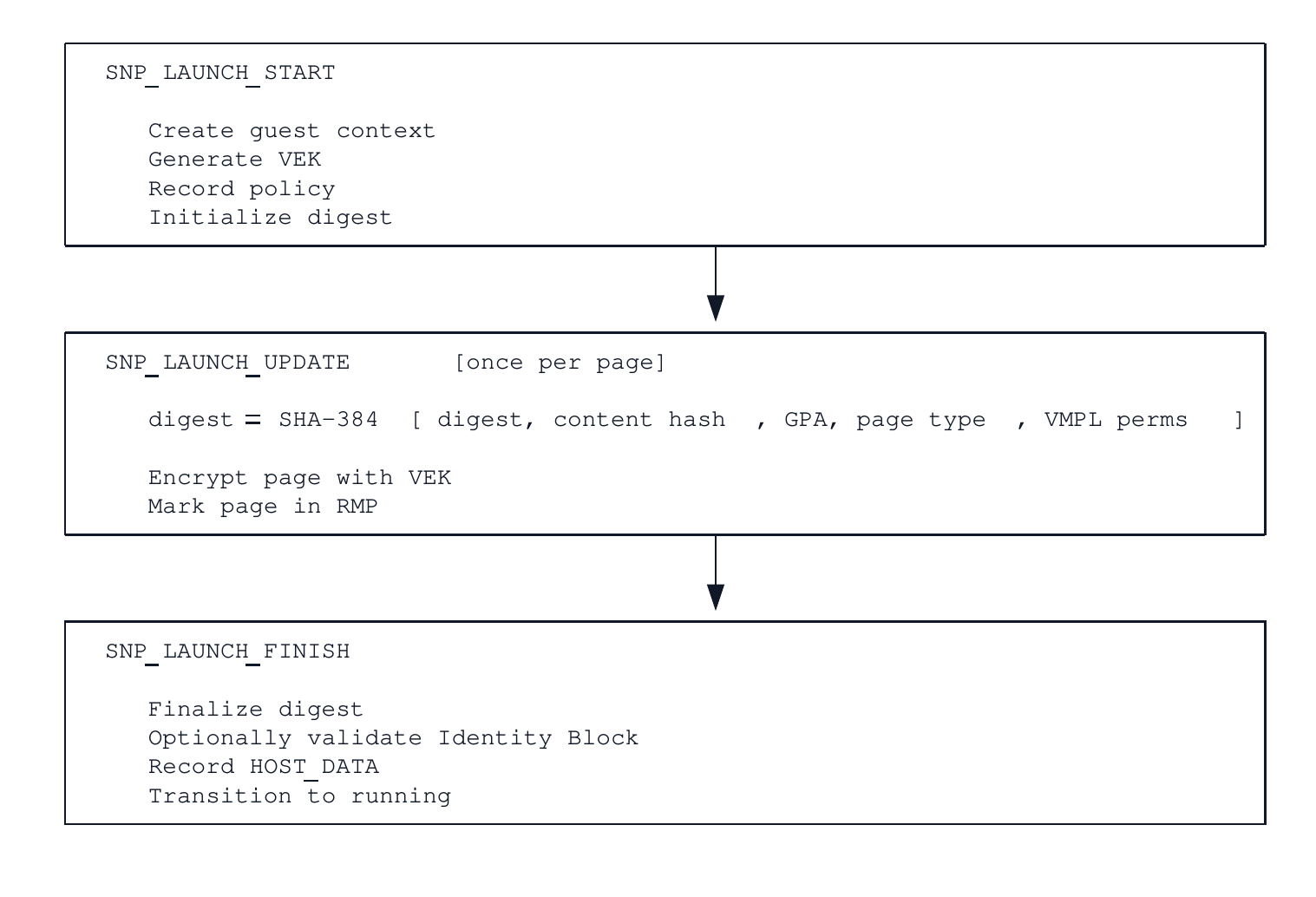}
\caption{The SNP launch sequence. SNP\_LAUNCH\_UPDATE is invoked once per page of initial memory; each call extends the digest by hashing the previous digest together with metadata describing the page (its content hash, GPA, type, and VMPL permissions). The final digest, finalized at SNP\_LAUNCH\_FINISH, is the value that appears as MEASUREMENT in subsequent attestation reports.}
\label{fig:23-snp-launch-sequence}
\end{figure}

The chain construction is what makes the digest meaningful. Each SNP\_LAUNCH\_UPDATE constructs a PAGE\_INFO structure containing the current digest, a SHA-384 of the page's contents (or zero for special types), the GPA where the page will be placed, the page type, and the VMPL permissions. The new digest is \texttt{SHA-384(PAGE\_INFO)}. Because each step incorporates the previous digest, the final value depends on the exact sequence of operations: loading the same code at different addresses produces a different digest, loading pages in a different order produces a different digest, and changing one byte in any page changes the final digest.

\begin{figure}[H]
\centering
\includegraphics[width=0.92\linewidth]{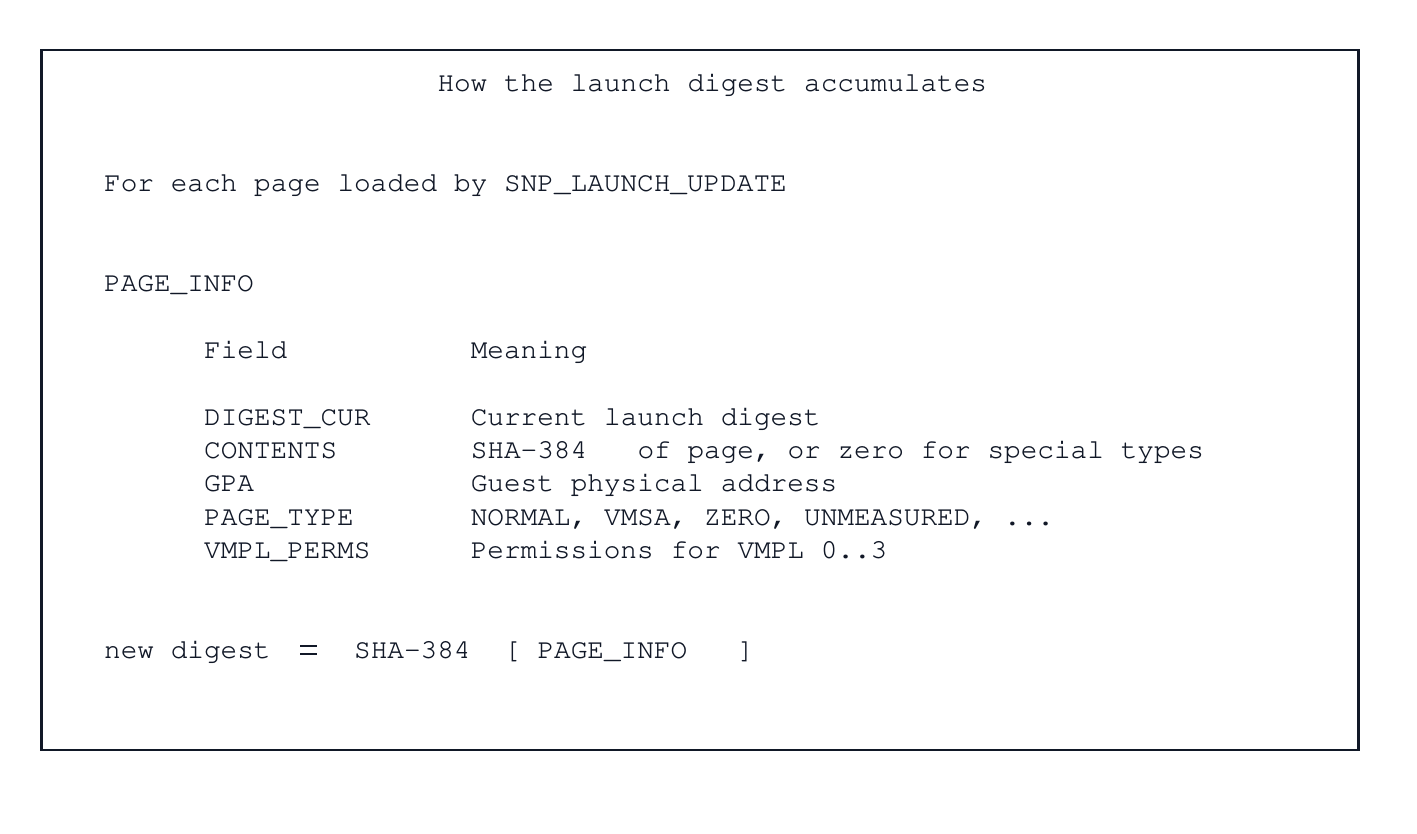}
\caption{How the launch digest accumulates. SNP\_LAUNCH\_UPDATE invokes one PAGE\_INFO computation per page; each invocation folds the previous digest, the page's content hash, its GPA, its type, and its VMPL permissions into a fresh SHA-384. The final value, finalized by SNP\_LAUNCH\_FINISH, is the MEASUREMENT field in subsequent attestation reports.}
\label{fig:24-launch-digest-chain}
\end{figure}

In a typical SNP launch with direct boot (passing kernel and initrd to QEMU through OVMF), the measured components are the OVMF firmware (the UEFI implementation), the Linux kernel, the initial ramdisk, the kernel command line, and the initial CPU state (the VMSA, which sets where execution starts). What is \emph{not} measured is anything loaded after boot: filesystems mounted from disk, code downloaded over the network, and runtime state changes.

Different page types are measured differently. Most pages are NORMAL pages whose contents are hashed in full. The VMSA page is measured as the initial CPU state, which determines where execution starts. ZERO pages are measured as zeros (the content is encrypted zeros, but the digest treats them as zero). UNMEASURED pages have only their GPA recorded, with content excluded from the hash; this is used for pages the guest will initialize itself. The SECRETS page is also unmeasured in content. The GPA is committed to in the digest, but the contents (the per-VMPL VMPCKs, written by the ASP after measurement) are not hashed. This way the measurement is reproducible (it does not depend on per-launch random key material) but the guest still knows where to find its keys. The CPUID page is similarly unmeasured in content. Its GPA is committed to, but the values (the hypervisor's CPUID responses, baked into the page at launch) are checked by the guest at runtime against the trusted launch image.

\subsection{The Attestation Report}
\label{subsec:the-attestation-report}

Once a guest is running, it can request an attestation report from the ASP. The request travels through the VMPCK-encrypted channel established by the secrets page (Section 5.5), so the hypervisor cannot read or modify it. The request includes 64 bytes of guest-supplied REPORT\_DATA and the VMPL of the requesting context. The ASP constructs the report and signs it.

\begin{figure}[H]
\centering
\includegraphics[width=0.92\linewidth]{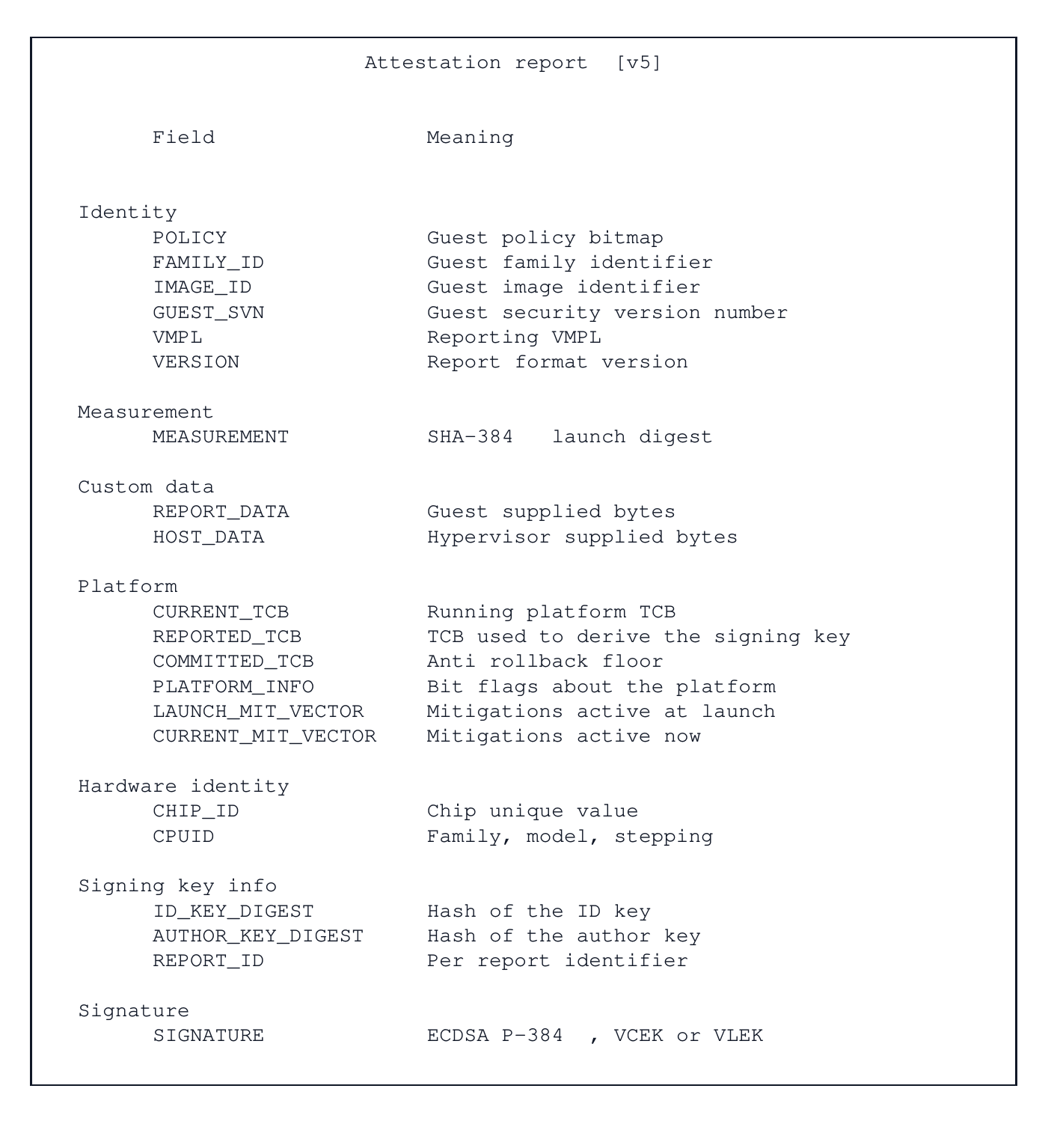}
\caption{The structure of an SNP attestation report (version 5). The report is a compact binary blob of approximately 1 KB, organized as identity, measurement, custom-data, platform, hardware-identification, signing-key, and signature sections, with the signature covering everything else.}
\label{fig:25-attestation-report}
\end{figure}

Two fields carry most of the feature visibility a verifier relies on. POLICY records the guest's launch-policy bits, among them DEBUG (which must be clear in production), a bit requiring ciphertext hiding, and PAGE\_SWAP\_DISABLE, which forbids the hypervisor's page-relocation commands; AMD added the last in response to attacks that abused page relocation as a ciphertext side channel \cite{ref14}. PLATFORM\_INFO records platform state: whether SMT is enabled, whether ciphertext hiding is active, whether the boot-time DRAM alias check completed, and whether Trusted I/O is enabled. A verifier can therefore confirm not just what launched, but the security posture of the environment it launched into.

Report version 5 also adds the two mitigation vectors, LAUNCH\_MIT\_VECTOR and CURRENT\_MIT\_VECTOR, which record which vulnerability mitigations were active at guest launch and which are active at report time. These let a verifier distinguish a platform that has a mitigation applied from one that merely runs newer firmware; the host can query and initiate mitigation verification through the SNP\_VERIFY\_MITIGATION firmware command \cite{ref3}.

\subsubsection{REPORT\_DATA}

The REPORT\_DATA field deserves attention because it is the principal mechanism by which attestation is bound to other cryptographic objects. The guest supplies these 64 bytes, and they are included in the signed report. Several patterns build on this.

To bind a TLS session to attestation, the guest generates an ephemeral TLS key pair, hashes the TLS public key, and supplies the hash as REPORT\_DATA when requesting a report. The verifier receives both the report and the TLS public key, checks that REPORT\_DATA matches the hash of the TLS key, and concludes that the TLS session terminates inside the attested guest. To bind freshness, the verifier supplies a random nonce, the guest includes the nonce in REPORT\_DATA, and the verifier checks the nonce on receipt. This proves the report is not a replay. To bind a manifest of additional claims, the guest produces a JSON or CBOR document describing claims it wants to commit to, hashes the document, and includes the hash in REPORT\_DATA. The verifier receives both the report and the manifest, checks the hash, and validates the manifest contents.

The 64-byte size is the constraint. For complex commitments, the standard pattern is to hash a larger document and place the hash in REPORT\_DATA.

\subsubsection{TCB versions}

The report contains four TCB-related fields, each serving a specific purpose. CURRENT\_TCB is the firmware/microcode versions actually running on the machine at the time of the report. REPORTED\_TCB is the TCB version that was used to derive the signing key; this can lag CURRENT\_TCB to allow an orderly key rotation after a firmware update. COMMITTED\_TCB sets a minimum below which the platform refuses to roll back. LAUNCH\_TCB records the TCB at the time the guest was created. On Milan and Genoa parts, each TCB value is a 64-bit field packing the ASP bootloader, ASP OS (TEE), SNP firmware, and CPU microcode versions into a single integer. Newer parts use a revised layout that adds a fifth component (the FMC field, carved from previously reserved bits), so verifiers must parse TCB values per CPU family rather than assuming a single universal layout \cite{ref3,ref25}.

The distinction between CURRENT\_TCB and REPORTED\_TCB is operational. After a firmware update, AMD's key distribution service needs time to make new VCEK certificates available, and during that window the platform can continue signing reports with the old TCB's key while the new firmware is already running. Verifiers enforce a minimum REPORTED\_TCB to reject signatures from firmware versions known to have vulnerabilities.

\subsection{The Key Hierarchy}
\label{subsec:the-key-hierarchy}

The signature on an attestation report has to chain back to something a verifier already trusts. AMD provides that anchor through a three-level key hierarchy ending at the VCEK (or, optionally, the VLEK).

\begin{figure}[H]
\centering
\includegraphics[width=0.92\linewidth]{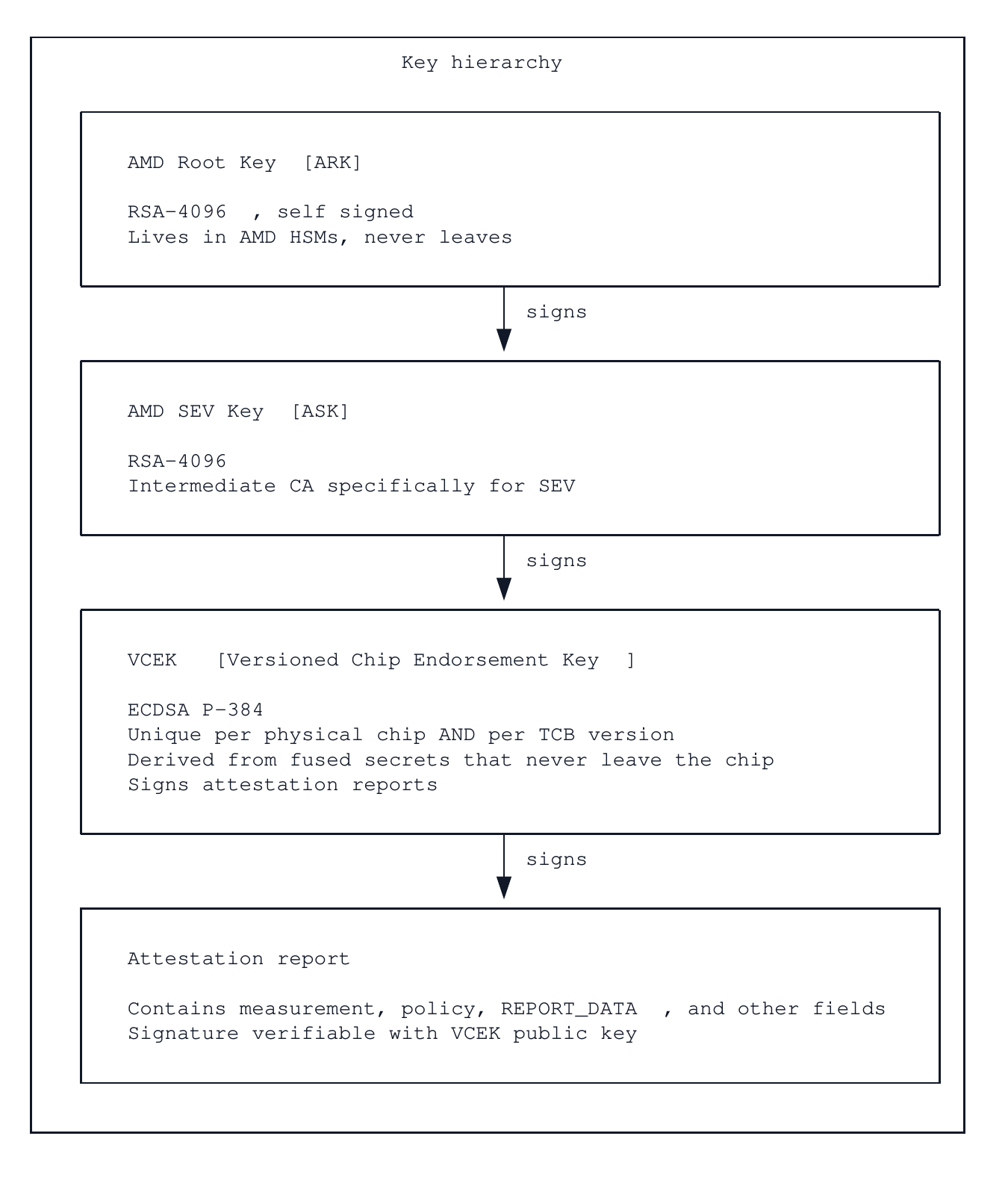}
\caption{The AMD attestation key hierarchy. The ARK is AMD's ultimate trust anchor, kept in HSMs and used only to sign the per-family ASK. The ASK signs per-chip VCEK certificates, which are unique to a physical chip and TCB version. The VCEK signs the attestation report.}
\label{fig:27-key-hierarchy}
\end{figure}

\subsubsection{The VCEK}

The VCEK has two properties that matter for verification. It is unique per physical chip, since each AMD processor has secrets fused into the silicon at manufacturing, and the VCEK is derived from those secrets, so two different chips produce different VCEKs even when running identical firmware. And it is unique per TCB version, because the derivation also incorporates firmware versions, so the same chip with different firmware produces a different VCEK. The derivation uses a one-way function, so knowing the VCEK for one TCB version does not let an attacker compute it for any other version. Only the chip's fused secrets, accessible to no software, can compute any VCEK.

That is the design intent, and it has not always held in practice. Researchers demonstrated a software-only extraction of the root VCEK seed on Milan parts by chaining a BootROM firmware-decryption flaw with missing fuse-write protections, which is enough to forge attestation reports at arbitrary TCB versions on affected parts \cite{ref16}. The attack relied on a chain of firmware vulnerabilities rather than on a weakness in the key derivation itself, and the initial vector was patched, but it is a concrete instance of the trust-anchor risk discussed in Section 2.5. The VCEK's strength is exactly the strength of the hardware and firmware surrounding it.

\subsubsection{VCEK vs VLEK}

The VCEK exposes which physical chip is running through the CHIP\_ID embedded in the report and the certificate. Some cloud providers consider this fingerprinting undesirable. AMD provides an alternative, the VLEK (Versioned Loaded Endorsement Key), which offers the same security guarantees with a slightly different trust model. AMD maintains VLEK seeds for enrolled cloud providers; the provider provisions the platform with a VLEK through a secure process. Reports can be signed with VLEK instead of VCEK, and VLEK certificates are issued by AMD but do not contain CHIP\_ID.

From the verifier's perspective, VCEK and VLEK provide equivalent assurance that the report was signed by genuine AMD hardware at a given TCB version, and they differ only in whether the specific chip is identifiable.

\subsection{Verifying a Report}
\label{subsec:verifying-a-report}

The verification flow has six steps.

\textbf{Step 1: receive the report.} The guest delivers its attestation report to the verifier, typically over an attestation-bound TLS session. The report is a binary blob of approximately 1 KB.

\textbf{Step 2: parse and extract metadata.} From the report, the verifier extracts CHIP\_ID, REPORTED\_TCB, and the CPUID family/model/stepping. These determine which certificates to fetch.

\textbf{Step 3: fetch certificates from AMD.} AMD's Key Distribution Service (KDS) at \texttt{kdsintf.amd.com} \cite{ref25} provides three resources. The certificate chain (\texttt{/vcek/v1/\{product\}/cert\_chain}) returns the ARK and ASK as a PEM bundle. The VCEK certificate (\texttt{/vcek/v1/\{product\}/\{chip\_id\}} with TCB SPL parameters) returns a DER-encoded certificate for the specific chip and TCB. The certificate revocation list (\texttt{/vcek/v1/\{product\}/crl}) returns the current CRL. For VLEK-signed reports, the analogous resources live under \texttt{/vlek/v1/}.

\textbf{Step 4: verify the certificate chain.} The verifier confirms that the ARK is self-signed (and that it matches a pinned copy of AMD's known ARK; fetching it from KDS without pinning would let a network attacker substitute a fake), that the ASK is signed by the ARK and within its validity period, and that the VCEK is signed by the ASK and within its validity period. The verifier also checks the VCEK's X.509 extensions for the correct TCB values and CHIP\_ID, and consults the CRL to ensure the VCEK has not been revoked.

\textbf{Step 5: verify the report signature.} Using the VCEK public key from the certificate, the verifier computes SHA-384 of the report bytes from offset 0x00 through 0x29F and verifies the ECDSA P-384 signature stored at offset 0x2A0. The signature format stores R and S components as zero-extended little-endian 72-byte values.

\textbf{Step 6: check report contents.} A valid signature only proves that the report came from genuine AMD hardware. The verifier still has to check that the report says what is expected. MEASUREMENT must match the expected launch digest. POLICY must be acceptable (DEBUG=0 in production deployments, other policy flags as required; current deployments should consider requiring PAGE\_SWAP\_DISABLE and, where the platform offers it, ciphertext hiding). PLATFORM\_INFO and the mitigation vectors should match the environment the deployment expects (alias check completed, ciphertext hiding active, required mitigations present). REPORTED\_TCB must meet the verifier's minimum, parsed per CPU family. REPORT\_DATA must contain the expected value (a TLS key hash, a fresh nonce, a manifest hash, depending on the binding the guest performed). VMPL must be the expected privilege level. Additional checks on FAMILY\_ID, IMAGE\_ID, ID\_KEY\_DIGEST, and other fields apply if the deployment uses them.

For verification to be meaningful, the verifier needs to know what measurement to expect. This requires the deployer to compute the expected launch digest in advance. The \texttt{sev-snp-measure} tool \cite{ref26} computes expected measurements given a description of the launch components (OVMF, kernel, initrd, kernel command line, vCPU count); confidential.ai's attestation-rs provides complementary attestation generation and verification tooling \cite{ref27}. For the computed value to match the actual launch, the build of each component must be byte-for-byte reproducible: a non-deterministic compiler, an unpinned dependency, or a timestamp embedded in a binary will all change the measurement.

\subsection{The Identity Block}
\label{subsec:the-identity-block}

The Identity Block (IDB) is an optional mechanism for asserting control over launch validation at the firmware level. Without an IDB, a malicious hypervisor could launch a modified VM image and the verifier would only detect this after the fact, when checking the attestation report. By that time the guest has already executed.

With an IDB, the guest owner pre-computes the expected measurement, signs a structure containing it, and provides this signed assertion to the hypervisor before launch. Concretely, before launch the guest owner builds the VM image and computes its expected launch digest, then constructs an Identity Block containing the expected digest, the expected policy, and metadata fields (FAMILY\_ID, IMAGE\_ID, GUEST\_SVN). The owner signs the IDB with an ID Key, and provides the IDB, signature, and public key to the hypervisor.

During launch, the ASP performs the normal measurement, and at SNP\_LAUNCH\_FINISH it checks that the computed measurement matches the IDB's expected digest and that the actual policy matches the IDB's expected policy, and that the signature is valid against the provided public key. If any check fails, the launch fails and the guest never runs. If all checks pass, the launch succeeds and the ID and author key digests appear in subsequent attestation reports.

The IDB also enables a signing-authority model. A verifier can confirm that the guest was launched with authorization from the holder of a specific key without needing to know the exact expected measurement, by validating ID\_KEY\_DIGEST and AUTHOR\_KEY\_DIGEST against an allow-list. This shifts the verifier's burden from tracking every measurement of every legitimate image to tracking the authorized signers.

\subsection{Attestation with an SVSM}
\label{subsec:attestation-with-an-svsm}

When an SVSM is present, attestation has additional considerations.

The SVSM is loaded and measured during launch, before the guest OS, so the launch digest reflects both. Verifiers must therefore know both the expected SVSM measurement and the expected guest firmware/kernel measurement. Different SVSM versions produce different launch digests.

Because the SVSM zeros VMPCK0 after startup (Section 5.5), only the SVSM can request reports tagged with VMPL=0. The guest OS, running at VMPL2 or VMPL3, uses VMPCK2 or VMPCK3 to request reports, and those reports are tagged with the corresponding VMPL. Deployments that need a VMPL0 attestation must obtain it from the SVSM.

The SVSM specification also defines an attestation extension that allows the SVSM to provide reports that include a manifest of the services it offers (vTPM, attestation proxy, and so on) and their properties. This lets a verifier check not only that the SVSM is the expected version but also that the SVSM is offering the expected services.

\subsection{What Attestation Does and Does Not Prove}
\label{subsec:what-attestation-does-and-does-not-prove}

The boundaries of what attestation guarantees are precise.

\begin{itemize}
  \item The report was generated by genuine AMD hardware running SEV-SNP.
  \item The hardware has a specific TCB version (firmware and microcode).
  \item The guest was launched with a specific measured set of bytes.
  \item The guest provided specific REPORT\_DATA at request time (whatever binding that REPORT\_DATA encodes).
  \item (If an IDB was used) a specific entity authorized this launch.
\end{itemize}

\begin{itemize}
  \item That the guest code is correct or secure. The measurement is an identity, not a behavioural assertion.
  \item That the guest has not been compromised after launch. Initial-state attestation says nothing about runtime state.
  \item That the hypervisor is behaving correctly outside the SEV-SNP boundary. Attestation says only that the boundary itself is active.
  \item That side channels are not leaking information.
  \item That the code does what the vendor claims. For closed-source code, attestation proves that a specific binary is running, not that the binary behaves as advertised.
\end{itemize}

\subsection{Verifier Responsibilities}
\label{subsec:verifier-responsibilities}

A correct verifier follows several disciplines that are not always made explicit in tool documentation. The ARK should be pinned, meaning a known-good copy bundled with the verifier and not fetched from KDS at every verification, since a network adversary could otherwise substitute a fake ARK. The CRL should be checked, because certificates can be revoked and a revoked VCEK should not be trusted. A minimum REPORTED\_TCB should be enforced, since old firmware may have known vulnerabilities. And REPORT\_DATA should actually be checked, since a report with a valid signature but the wrong REPORT\_DATA may be a replay or a misdirection.

The attacks that remain possible against a correctly verified report are bounded. The hypervisor can refuse to let the guest request a report, or refuse to deliver it, which amounts to denial of service. It can also roll back within the committed TCB, since COMMITTED\_TCB sets a floor and the hypervisor might run an older but still-permitted version. If the guest can be exploited before it requests attestation, the attacker could make the guest commit to a key the attacker controls. And attestation reports the initial state, so an exploit that occurs after the report is generated will not be reflected in any prior report.

These limits are inherent to attestation as a primitive.

\section{Summary}
\label{sec:summary}

The SEV-SNP design is a coherent answer to a specific question: how does one execute a virtual machine on infrastructure controlled by an untrusted party while still preserving the confidentiality and integrity of what the virtual machine processes?

The answer comes in layers. The threat model (Section 2) defines the boundary. Confidentiality and integrity of in-memory data and registers are guaranteed against a software-level adversary with full hypervisor control. Availability, side channels, software bugs, and invasive physical attacks are explicitly out of scope. The hardware foundations (Section 3) supply the primitives. The ASP holds keys that x86 software cannot read, the on-die memory controller encrypts data crossing the package boundary, and nested paging together with ASID-based key selection let multiple confidential VMs coexist. Current processors extend that foundation (ciphertext hiding, hardware-enforced interrupt filtering, RMP segmentation, secure timekeeping, and trusted I/O), with each addition individually enabled at the platform or guest level. The integrity layer (Section 4) adds the Reverse Map Table, an inline check on every relevant memory access that ensures a guest's view of memory is consistent with what the guest itself wrote, regardless of what the hypervisor has done to the underlying mapping. The privilege and communication layer (Section 5) protects register state in the encrypted VMSA, allows internal privilege separation through VMPLs and an SVSM, and gives the guest explicit control over what crosses the trust boundary through the GHCB protocol. Attestation (Section 6) closes the system by binding a hardware-signed assertion of the guest's identity to a key chained to AMD's silicon, so a remote verifier can confirm independently what is running.

The trust assumptions of a SEV-SNP deployment are sharply different from those of a conventional cloud deployment. A conventional deployment trusts the cloud provider's entire software stack: hypervisor, host OS, management plane, orchestration, monitoring, and the operators of all of those. A SEV-SNP deployment trusts AMD's manufacturing, AMD's key management, AMD's ASP firmware, the cryptographic primitives the design relies on, and the physical hardware host's commitment not to mount invasive physical attacks. This is not zero trust. It is a smaller, more concentrated, and more legible trust surface than the alternative. The assumptions are explicit, public, and bounded. If AMD's silicon or firmware is compromised, the guarantees fail, but the failure mode is a single well-defined event rather than the diffuse risk profile of a conventional deployment, where any one of many parties might misbehave.

The mechanisms described in this paper are concrete implementation details of one TEE architecture, but the conceptual machinery they instantiate is more general. A hardware root of trust signs a measurement. A guest cooperates with an untrusted hypervisor through an explicit, narrow channel. Memory is encrypted on the way out of the package and integrity-checked on the way in. Trust extends from a launch-time measurement to runtime code through measured chains (the kernel measures the rootfs root hash, dm-verity verifies blocks, the vTPM seals keys against system state). These same patterns recur across the broader landscape of confidential computing technologies. SEV-SNP is one particular point in that design space; understanding it in detail is a good basis for understanding the others.

\end{document}